\documentclass[12pt]{article}
\usepackage{amsmath}
\usepackage[margin=2.5cm]{geometry}

\usepackage{graphicx}
\usepackage{subcaption}
\usepackage{csquotes}
\usepackage[auth-sc, affil-sl]{authblk}

\usepackage{titlesec}

\titleformat{\section}[block] {\normalfont\sffamily}{\textbf{\thesection}}{.5em}{\bfseries}

\titleformat{\subsection}[block] {\normalfont\sffamily} {\textbf{\thesubsection}}{.5em}{\bfseries}

\titleformat{\subsubsection}[block] {\normalfont\sffamily} {\textbf{\thesubsubsection}}{.5em}{\bfseries}

\newcommand{\pd}[2]{\ensuremath{\frac{\partial #1}{\partial #2}}}

\usepackage{xcolor}

\usepackage{epsfig}
\usepackage{amsfonts}
\usepackage{fancyhdr}
\usepackage{cite}
\usepackage{hyperref}
\usepackage[nolists, nomarkers]{endfloat}
\usepackage{setspace}
\newcommand{\runningtitle}{Fifty years of Schallamach waves} % NOTE: Remember to change this line!
\newcommand{\authorname}{\small K. Viswanathan \& S. Chandrasekar} % NOTE: Remember to change this to your name!
\newcommand{\mt}[1]{{\color{black}{#1}}}
\newcommand{\sV}{\ensuremath{V_0}}

\title{\Large \sffamily \textbf{Fifty years of Schallamach waves: From rubber friction to nanoscale fracture}}
\author[1]{\normalsize Koushik Viswanathan\thanks{Corresponding author: koushik@iisc.ac.in}}
\author[2]{\normalsize Srinivasan Chandrasekar}
\affil[1]{Department of Mechanical Engineering, Indian Institute of Science, Bengaluru, India}
\affil[2]{Center for Materials Processing and Tribology, Purdue University, USA}
\date{\small \vspace{-1em} May 30, 2022\vspace{-2em}}
\begin{document}
\maketitle
\thispagestyle{plain}
\noindent\rule{\textwidth}{0.5pt}
\vspace{-3em}
\begin{abstract}
  The question of how soft polymers slide against hard surfaces is of significant scientific interest, given its practical implications. Specifically, such sytems commonly show interesting stick--slip dynamics, wherein the interface moves intermittently despite uniform remote loading. \mt{The year 2021 marked the 50$^{th}$ anniversary of the publication of a seminal paper by Adolf Schallamach (\emph{Wear}, 1971)} that first revealed an intimate link between stick--slip and moving detachment waves, now called Schallamach waves. We place Schallamach's results in a broader context and review subsequent investigations of stick--slip, before discussing recent observations of solitary Schallamach waves. This variant is not observable in standard contacts so that a special cylindrical contact must be used to quantify its properties. The latter configuration also reveals the occurrence of a dual wave---the so-called separation pulse---that propagates in a direction opposite to Schallamach waves. We show how the dual wave and other, more general, Schallamach-type waves can be described using continuum theory, and provide pointers for future research. In the process, fundamental analogues of Schallamach-type waves emerge in nanoscale mechanics and interface fracture. The result is an on-going application of lessons learnt from Schallamach-type waves to better understand these latter phenomena. 

\end{abstract}
\vspace{-2em}
\noindent\rule{\textwidth}{0.5pt}

\section{Introduction}

The question \lq \emph{How does rubber slide?}\rq\ marks the title of a landmark paper by Adolf Schallamach published 50 years ago \cite{Schallamach_Wear_1971}. This work investigated the basic mechanisms of sliding friction when a soft compliant object, such as rubber, is slid against a much harder and practically rigid object, such as glass. The question appears to be a humble one, and the 1971 paper was one of the many pinnacles of a broad research programme on rubber friction at the \mt{Natural Rubber Producers' Research Association} in the United Kingdom. Testament to its success is the fact that this seemingly simple question has thrown up a rich variety of dynamic behaviour in response, spanning length and time scales, and occurring in completely unexpected situations \cite{Barquins_Wear_1993}. 

In order to fully appreciate this complexity, consider a simple configuration comprising two semi-infinte solids that are in contact along the $xy$ plane and under a remotely applied normal force $N$ in the $z$-direction. We assume that one solid is elastic $(z\leq 0)$ and the other rigid $(z>0)$. If, at time $t=0$, the elastic solid is subject to a constant remote velocity $\sV$ parallel to its contact plane\footnote{Note that we will henceforth use the same notation for the remote sliding velocity.}, how does motion at the interface occur? Note that we have posed the problem via an applied velocity \emph{in lieu} of an applied remote shear force to ensure that the solids will eventually slide against each other. Elementary mechanics tells us that if \sV\ is constant, the shear force transmitted at the interface will continuously increase from $t=0$ until it reaches the static friction threshold $F_s = \mu N$, determined by a coefficient of friction $\mu$, presumed known for this pair of solids, and the normal force $N$. Beyond this threshold, uniform sliding should ensue at constant interface shear force $F_d$, the so-called \lq dynamic\rq\ friction force. This is the standard Coulomb model for friction \cite{BowdenTabor_Friction_1973}. 

Schallamach's work dramatically exposed the shortcomings of this line of thought. Firstly, it showed that the interface itself does not often move uniformly but instead only via the propogation of localized wave-like detachment zones---now aptly named Schallamach waves in his honour---that traverse the entire length of the contact interface. Detachment waves of this nature can best be described by analogy with the motion of a ruck in a carpet \cite{VellaETAL_PhysRevLett_2009}: if a carpet is to be moved by distance $\Delta x$, we could either simply translate the entire carpet surface at once by $\Delta x$, or create a localized slip zone by buckling---causing slip $\Delta x$---that then propagates along the carpet, progressively causing it to slip. This buckling and detachment type mechanism is reminiscent of that proposed by Bragg for explaining how dislocations mediate slip in crystals \cite{PhillipsThomas_BraggLegacy_1990}. 

The most distinguishing properties of Schallamach waves are that they cause local interface slip, and that they propagate significantly more slowly than any known elastic waves, e.g., Rayleigh and Stoneley waves. It is only now being appreciated that these waves are part of a larger class of propagating detachment fronts that arise not only in frictional interactions, but also in nanoscale fracture \cite{GerdeMarder_Nature_2001}, seismology \cite{Scholz_Nature_1998, Heaton_PhysEarthPlanInt_1990} and even biological locomotion \cite{GrayLissman_JExptBio_1938, Trueman_SoftBodiedLocomotion_1975, LaiETAL_JExptBio_2010}.

Schallamach's work also established the importance of elastic deformation and adhesion in determining the frictional response of an interface. The aforementioned Coulomb friction model for the onset of steady sliding applies only to perfectly rigid solids. When one of the bodies is allowed to deform, the interface dynamics can change significantly \cite{BurridgeKnopoff_1967, CartwrightETAL_PhysRevLett_1997, ViswanathanSundaram_Wear_2017}. Likewise, when the two bodies have significant interface adhesion, reattachment is easily effected following local detachment, leading to localized pockets that can then move and cause slip \cite{Kendall_Nature_1976}.

In this article, we revisit various facets of Schallamach waves, fifty years after they were first reported. The first part of this manuscript presents a topical review of Schallamach's original contributions and their continued impact (Sec.~\ref{sec:SchallamachPaper}). Our experimental results are then presented in Sec.~\ref{sec:exptl}. The solitary Schallamach wave is introduced, along with another detachment wave that moves in the opposite direction---the \lq dual\rq\ of the Schallamach wave, also termed the separation pulse. By condensing and abstracting the basic features of the Schallamach wave and its dual, we show how more general traveling waves can be grouped together as \lq Schallamach-type\rq\ waves. This potentially broadens the applicability of lessons learned from studying waves in one situation to an entirely different domain. Section~\ref{sec:theory} presents results from prior and on-going theoretical investigations of Schallamach-type waves at interfaces. Experimental observations have led to the emergence of new theoretical ideas, even within the centuries-old framework of continuum elasticity. We conclude our manuscript with a brief discussion and an outlook for future work in this domain.

\section{Schallamach's experiments and their primary significance}
\label{sec:SchallamachPaper}
\begin{figure}[ht!]
  \centering
  \includegraphics[width=0.3\textwidth]{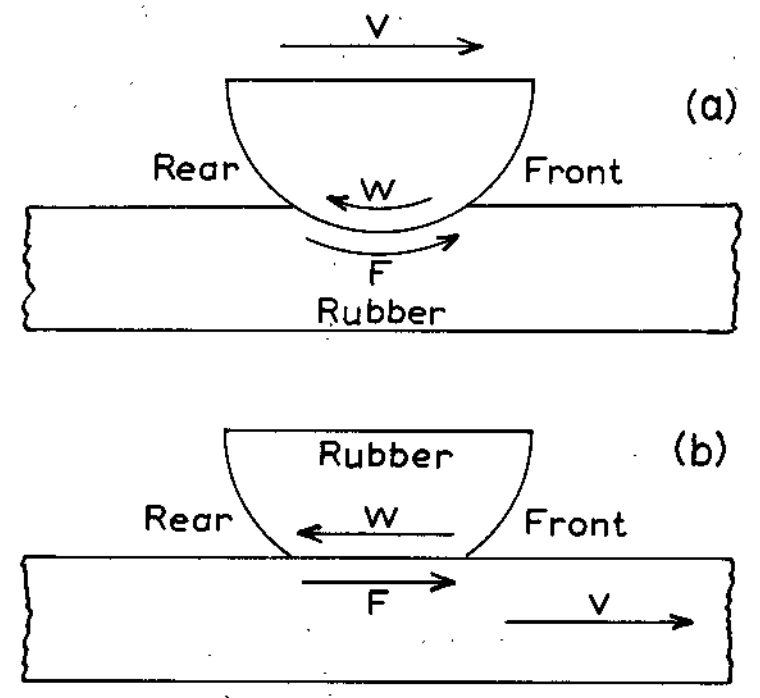}\quad\quad
  \includegraphics[width=0.6\textwidth]{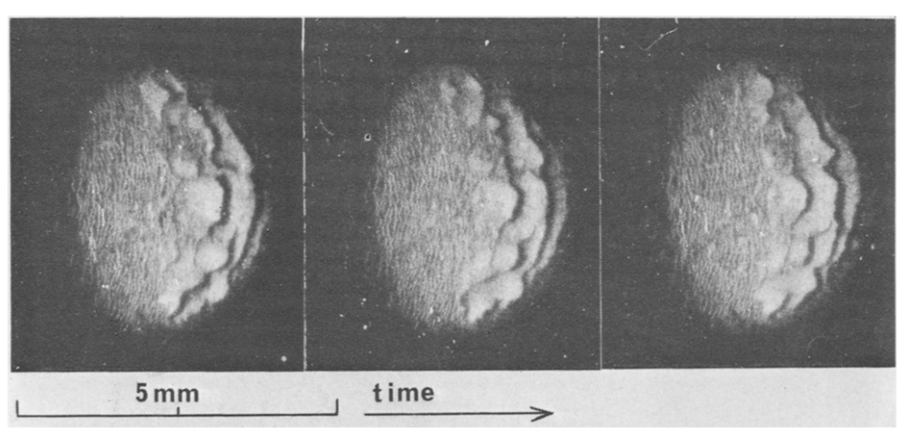}
  \caption{Panels showing Schallamach waves in an adhesive contact. (Left) The two different experimental configurations used by Schallamach comprise of a spherical glass indenter sliding against a rubber flat (a) and a spherical rubber indenter sliding against a glass flat (b). Note the direction of friction force $F$ vis-\'a-vis wave propagation direction $w$. (Rigt) Three frames showing the propagation of multiple Schallamach waves, sliding direction is as shown in (a) on the left. Adapted from Ref.~\cite{Schallamach_Wear_1971}}
  \label{fig:orig-frames}
\end{figure}

Schallamach's original observation of detachment waves were made at a rubber--glass interface, under two complementary geometries---a glass hemisphere slid at constant speed $V_0$ against a rubber flat, and a rubber hemisphere slid against a glass flat. In either case, when the sliding velocity was low but above a critical value $V_0 > V_0^C$, the interface did not move uniformly at speed $V_0$, but instead slipped intermittently, see Fig.~\ref{fig:orig-frames} reproduced from Schallamach's paper \cite{Schallamach_Wear_1971}. The left panel in this figure shows a schematic of the two sliding geometries. The images in the right panel correspond to experiments with a glass indenter/rubber flat and were obtained by directly observing the interface through the transparent glass hemisphere and lit by a backlight source. The view here is along the normal of the plane flat surface. The contact zone is grey and the non-contacting zones appear darker due to the lighting. 

These frames show several creases or folds traversing from right to left within the contact zone. Each crease is a single Schallamach wave and is comprised of a local region of detachment so that it appears dark in the image. The frames clearly show multiple waves within the contact zone at any time instant. Well defined individual waves, with larger detachment regions, occur on the right side of the contact, and become increasingly indistinguishable towards the left. The result is an intermittent motion of the interface---each crease or wave causes a small amount of slip, and a point on the interface remains stationary until a wave traverses it. This was deduced by Schallamach as follows:
\begin{displayquote}
  \textsl{Dust particles fortuitously caught in the interface were seen to be buffetted by successive waves, and to remain motionless in the intervals\ldots A (single) wave of detachment is a fold in the rubber surface with presumably large strains around its three curved parts. To propagate the wave, energy must be supplied to cover the energy losses incurred when this strain configuration travels along the contact.}
\end{displayquote}
Strong interface adhesion between the two solids was necessary for waves to occur. Adhesion hysteresis was identified as a source of energy loss, in addition to that needed for changing the strain configuration in the vicinity of the propagating wave. Schallamach also noted that the wave propagation direction $w$ in the schematic of Fig.~\ref{fig:orig-frames} is always against the direction of friction force $F$ on the rubber in either contact configuration (hemisphere or flat). This direction of propagation has important significance that will be elaborated upon in later sections.

Finally, Schallamach also measured propagation speeds of individual waves, along with their frequency (no. of waves per second within the contact) and found that the latter was nearly proportional to the imposed $\sV$. Further, the wave speed he measured ($\sim 0.01$ m/s) was much lesser than typical elastic wave speeds ($\sim  100$ m/s). This is after accounting for the fact that the wave speed was \lq \textsl{more difficult to assess}\rq . Therefore, the possibility that these waves of detachment were associated with any known elastic surface waves appeared remote at first sight. 

\section{Experimental observations of Schallamach waves}
\label{sec:exptl}

\mt{The most common configuration for observing Schallamach waves is the spherical geometry described in the previous section. We first summarize experimental observations using such a spherical indenter geometry and use them to explain why Schallamach waves must nucleate in the first place. This is followed by a description of a more recent configuration presented by Viswanathan \emph{et al.} \cite{ViswanathanETAL_PhysRevE_2015} that uses a cylindrical contact geometry to isolate single waves without intervening edge effects.} 

%the same spherical contact configuration studied in Schallamach's work, in order to obtain more quantitative information about their nucleation mechanisms. Following this, we introduce a very different configuration---involving cylindrical contact---specifically to isolate single (or solitary) waves and study their properties. Such solitary waves show several features that suggest analogies with nanoscale deformation phenomena, which are further explored below.

\mt{\subsection{Spherical contacts: Why do Schallamach waves nucleate?}}

\begin{figure}[ht!]
  \centering
  \includegraphics[width=0.6\textwidth]{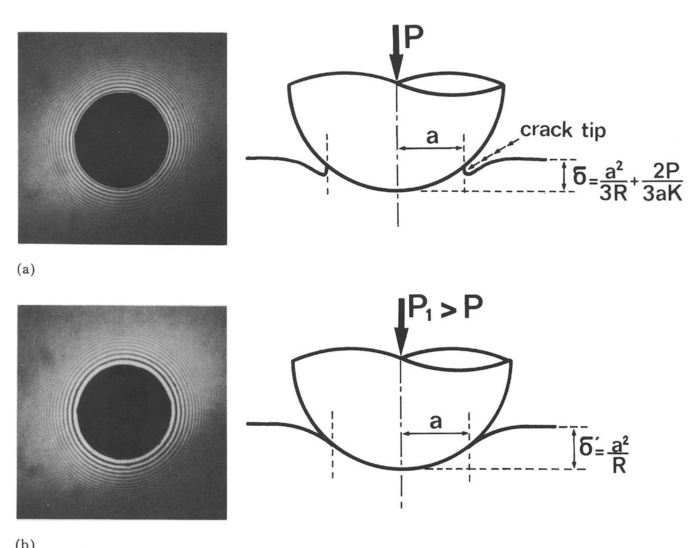}
  \caption{Schematic comparing standard Hertzian non-adhesive contact (bottom) and JKR-type adhesive contact (top). In the former, the rubber surface is nearly tangential to the indenter profile. In the latter, it is normal to the surface, due to intermolecular attraction. Adapted from Ref.~\cite{Barquins_Wear_1985}.}
  \label{fig:barquins-nuc}
\end{figure}

Before we discuss the formation of Schallamach waves, it is first important to consider the mechanics of the spherical contact in some detail. Here and henceforth, we assume that the indenting object is rigid and the indented object is elastically deformable. The latter is also assumed to be semi-infinite (half-space) as an idealization. This situation is schematically demonstrated in Fig.~\ref{fig:barquins-nuc}, reproduced from an early review article on Schallamach waves \cite{Barquins_Wear_1985} . When a rigid hemisphere of radius $R$ is pressed against an elastic surface with a load $P$, the latter deforms, resulting in a contact radius of size $a(P)$. It is well known that the depth of penetration $\delta = a^2/R$ depends on the load $P$ via the Hertz contact relations \cite{Johnson_ContactMechanics}. Additionally, the edges of the contact zone are tangential to the spherical surface (Fig.~\ref{fig:barquins-nuc}(b)).

In the presence of interface adhesion, the contact zone is very different \cite{JohnsonETAL_ProcRoySocA_1971}, see Fig.~\ref{fig:barquins-nuc}(a). Notably, the edge of the contact zone is now normal to the spherical indenter and its geometry resembles that of a crack-tip \cite{MaugisBarquins_1980}. This has two important consequences: firstly, the contact radius $a$ is now larger for the same force $P$, so that $\delta$ is also larger. Secondly, and more importantly, the surface immediately outside the contact zone is depressed and surrounded by a region that is raised. This is because rubber is essentially incompressible so that a compressive strain in one region must be compensated by comparable tensile strain nearby.

\begin{figure}[ht!]
  \centering
  \includegraphics[width=0.9\textwidth]{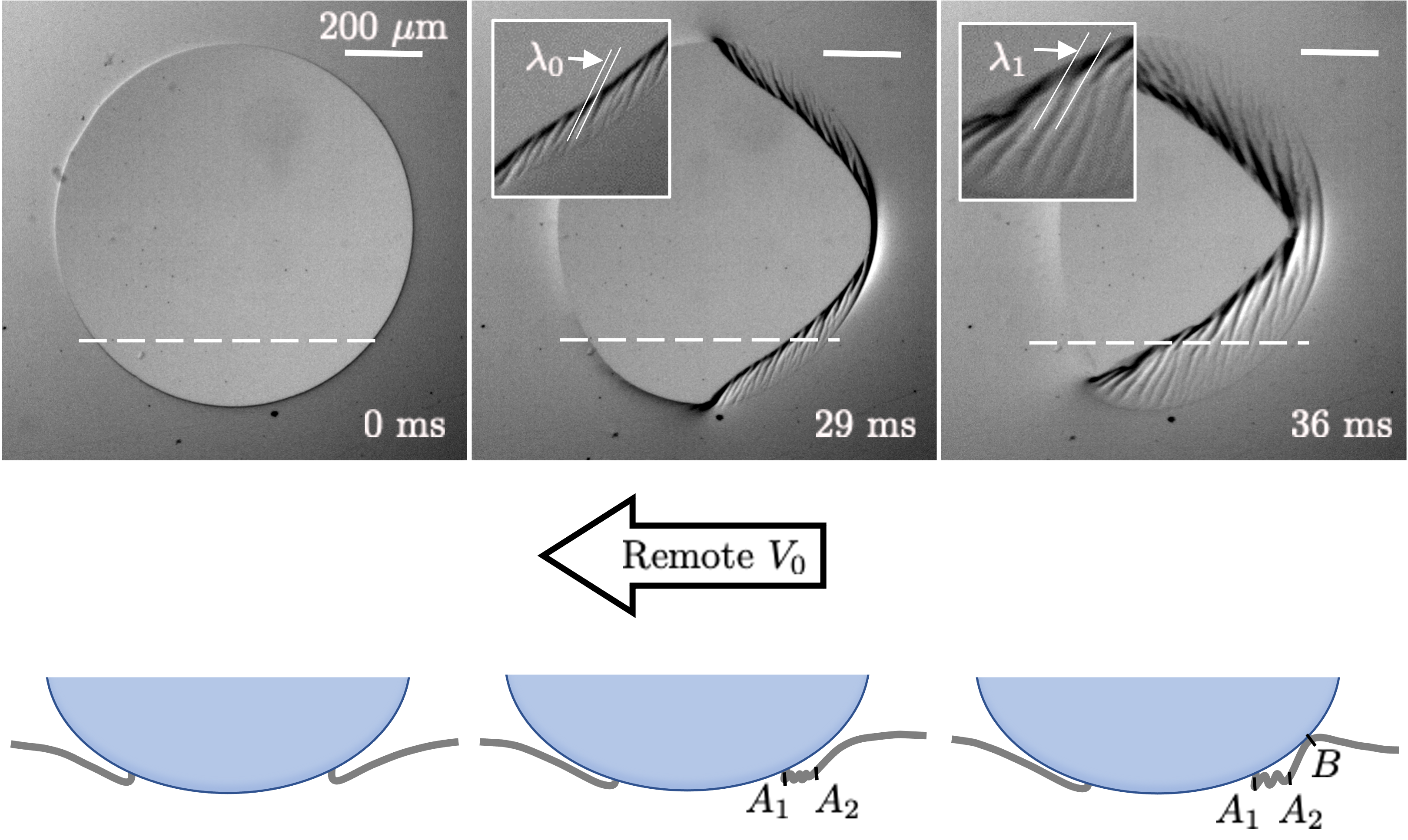}
  \caption{Nucleation of a single Schallamach wave within the spherical contact geometry. The elastic polymer is slid at constant remote velocity $V_0$ to the left and initially remains stationary at the interface (frame 1). Buckling results in a pronounced bulge on the free surface, with a corresponding surface depression between points $A_1$ and $A_2$ (frame 2). The bulge eventually grows enough to reattach to the indenter at point $B$ (frame 3). Wrinkles accompanying a nucleating Schallamach wave show an initial wavelength $\lambda_0$ (inset, frame 2) which doubles to $\lambda_1 \simeq 2 \lambda_0$ (inset, frame 3) just at the onset of wave propagation. The inset locations are marked by white dashed lines in the corresponding frames. }
  \label{fig:wrinkles}
\end{figure}

With this background, the reason why a Schallamach wave must nucleate can be clearly stated, see Fig.~\ref{fig:wrinkles}, adapted from Ref.~\cite{ViswanathanETAL_PhysRevE_2015}. Here, the indenter is a glass lens and the elastic substrate is polydimethylsiloxane (PDMS). Since the indenter and the substrate form an adhesive contact, the edge of this contact zone forms a crack-like interface, see frame 1 and corresponding side-view schematic below. Upon application of a remote $\sV$ to the elastic polymer, the tangential force $T$ increases monotonically while adhesion causes the interface to remain stationary. However, at a critical load, a region immediately ahead of the contact undergoes surface buckling causing a bulge to form on the surface and a corresponding depression between the points $A_1$ and $A_2$ (frame 2). As $T$ is increased further, the bulge grows and eventually makes contact with the indenter at a second location $B$. This detached region between the two contact locations $A_1$ and $B$ constitutes a nucleated Schallamach wave that then begins to propagate through the interface. Since $\sV$ is maintained constant, wave propagation is accompanied by a reduction in $T$ and the entire process repeats again to nucleate another wave. It is therefore clear from this sequence that each wave is nucleated by a surface buckling instability. The result is intermittent motion of the interface---commonly termed stick--slip---with some parts remaining stationary and others (moving wave regions) slipping.

An analogous mechanism was also postulated by Schallamach:
\begin{displayquote}
  \textsl{
    On applying a steadily increasing tangential stress, sliding begins when the compressive stress in front of the contact area has become great enough for buckling}
\end{displayquote}
It is evident from these observations that adhesion plays a key role not only in sustaining wave motion but also in wave nucleation, both by changing the contact geometry (making buckling more conducive) and by effecting reattachment. \mt{The existence of the critical (buckling) load explains why Schallamach waves are only observed after a threshold velocity $V_0^C$ is crossed. Note that this type of constant velocity boundary loading has historically been the most common configuration in friction experiments, as opposed to constant shear stress or force. The reason for this is that the transition from static to dynamic motion is best observed in this situation \cite{Rabinowicz_SciAm_1956, Ruina_JGeophysRes_1983}.}

Compression-driven buckling has an additional side effect---the occurrence of secondary features such as wrinkles on the detached surface \cite{KoudineBarquins_JAdhSciTech_1996}. A half-space subject to compression inevitably shows wrinkles, folds and creases at the surface \cite{Biot_ApplSciRes_1963, Groenewold_PhysicaA_2001, BrauETAL_SoftMatter_2013}. Sample wrinkles are shown in the insets to frames 2, 3 of Fig.~\ref{fig:wrinkles}. The wrinkle pattern that forms first has an inherent wavelength $\lambda_0$ that is determined by the depth of the strained surface layer (inset, frame 2). This wavelength doubles as the tangential load $T$ is increased so that $\lambda_1 \sim 2\lambda_0$ (inset, frame 3). As the detached zone begins to propagate, wrinkles result in imperfect readhesion in its wake. Consequently, the size of these pockets of imperfect readhesion is nearly equal to the final wrinkle wavelength $\lambda_1$. 

\mt{The buckling-induced mechanism for Schallamach waves is qualitatively similar in most situations, even, in the presence of a lubricant \cite{Cushman_Nature_1990}. However, the presence of an interfacial liquid film changes the nature of the resulting stick--slip process quantitatively \cite{WuBavouzet_2007}. Firstly, the threshold $V_0^C$ is lowered, perhaps because the increased contact pressure due to fluid penetration inside the contact enhances the propensity for buckling. Note that this possibiliy does not occur if the indenter is rigid (as in Fig.~\ref{fig:barquins-nuc}) since then the contact remains dry with minimal fluid ingress. Secondly, at much higher velocities, the picture appears to be fundamentally altered in liquids such as water-glycerol that can wet and dewet from rubber surfaces \cite{MartinETAL_PhysRevE_2002}. Now sticking and slipping occurs not due to repeated buckling instabilities but due to a wetting-squeezing-dewetting process. These \lq dewetting waves\rq\ are not the same as Schallamach waves and will not be considered further in this article. It is therefore safe to conclude that surface buckling is well established as the underlying cause for the nucleation of Schallamach waves \cite{BestETAL_Wear_1981,  RandCrosby_ApplPhysLett_2006, FukahoriETAL_Wear_2010}.}

{\subsection{\mt{Cylindrical \emph{vs.} spherical geometry}}
\label{subsec:exptl}
%\mt{a hemi-spherical glass lens (Edmund Optics) is used as the indenter and an elastic polymer (polydimethylsiloxane, Dow Corning) as the elastic substrate. Constant remote velocity $\sV$ is applied via a ball-screw driven linear slide actuated by a servo motor. The first consists of a spherical contact geometry, recreating Schallamach's original configuration (Sec.~\ref{sec:SchallamachPaper}).
\begin{figure}[ht!]
  \centering
  \includegraphics[width=\textwidth]{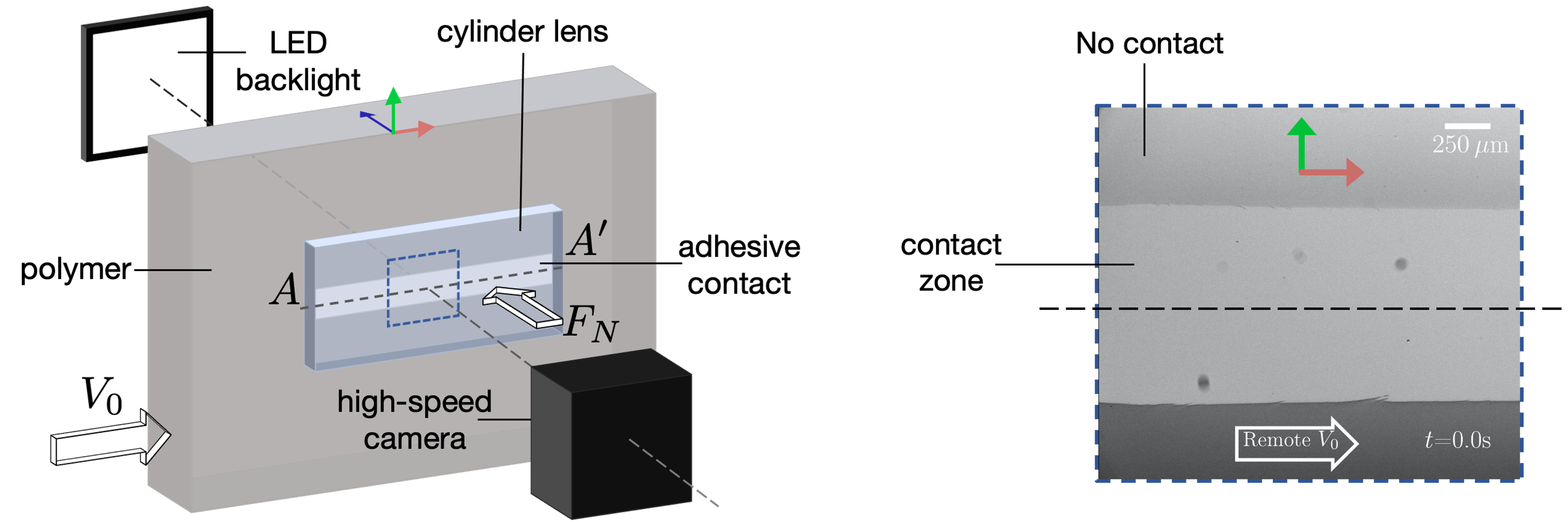}
  \caption{Experimental schematic showing a typical cylindrical lens in adhesive contact leading to the formation of an \lq adhesive\rq\ channel. The schematic on the left shows a typical configuration with a backlit source and remote applied motion $V_0$. A sample image showing the contact zone (light grey) and detached zones (dark grey) is presented in the right panel. \mt{Note that the red, green and blue arrows represent the $x, y,z$ axes, respectively.}}
  \label{fig:schematic-adhesive}
\end{figure}

%The second experimental configuration we use comprises a cylindrical geometry, see Fig.~\ref{fig:schematic-adhesive}.
%In order to study the properties of individual Schallamach waves unencumbered by contact end-effects, one must ideally use a plane 2D contact geometry. However, this leads to practical issues such as contact misalignment and occurrence of multiple nucleation sites. A trade-off is to instead use a cylindrical indenter with large radius, small contact force and long length \cite{ViswanathanETAL_SoftMatter_2016_1}. The resulting contact then resembles a long aspect ratio \lq adhesive channel\rq\ within which single Schallamach wave nucleation can easily occur.

\mt{In contrast to the most commonly used spherical contact configuration discussed above, a second cylindrical contact geometry is now considered, see Fig.~\ref{fig:schematic-adhesive}. This geometry is inspired by the need for observing Schallamach wave propagation in a nearly planar interface without any edge effects. A fully flat configuration has its own challenges, especially with respect to aligning the indenter and the substrate \cite{YamaguchiETAL_JGeoPhysRes_2011}. The use of a large radius cylindrical indenter compenstates for this since, at low load, contact is first initiated along the cylinder axis (line contact) \cite{ViswanathanETAL_SoftMatter_2016_1}.}   

\mt{The contact geometry we describe here is adapted from Ref.~\cite{ViswanathanETAL_PhysRevE_2015}.} A cylindrical plano-convex lens (\mt{Edmund Optics, radius 16.25 mm, length 25 mm}) is used as the indenter and the substrate is again polydimethylsiloxane (as in Fig.~\ref{fig:wrinkles}). \mt{The PDMS samples were produced mixing a base (vinyl-terminated polydimethylsiloxane) with a curing agent (methylhydrosiloxane-dimethylsiloxane copolymer) in the ratio 10:1 by weight. The resulting mixture was cured for 12 hours at 60$^\circ$C in the form of slab-like specimens \cite{ViswanathanETAL_PhysRevE_2015}. The Poisson's ratio and Young's modulus of the final samples were 0.46 and 1 MPa, respectively.} Constant remote $\sV$ is effected via the same ball-screw mechanism as before and the contact is imaged using a high-speed camera (pco Dimax) attached to a microscope system. The right hand panel in Fig.~\ref{fig:schematic-adhesive} presents a typical image obtained by viewing the backlit contact through the transparent cylindrical indenter. The lighter (darker) regions correspond to interface contact (detachment) just as before.

In both cases, the polymer sample had dimensions along the $x,y,z$ axes (Fig.~\ref{fig:schematic-adhesive}) of 70 mm $\times$ 22 mm $\times$ 25 mm, respectively, with the long dimension representing the sliding direction. Prior to any experiment, the PDMS and lens are brought into contact along the $x$-axis and maintained for a fixed time $t=60$ s to standardize any possible contact aging effects. Contact alignment is also checked during this period by translating the camera/light source along the $x$-axis. The PDMS is then slid relative to the lens using the linear stage at constant speeds $\sV = $10 $\mu$m/s to 20 mm/s, for a distance of 30 mm. Concurrent with the \emph{in situ} imaging, sliding forces were measured using a piezoelectric dynamometer (Kistler).

\subsection{The solitary Schallamach wave and its opposite moving dual}
\label{subsec:solitarySchallamach}

One of the primary problems with the standard spherical indenter geometry used for studying Schallamach waves is its inability to isolate single waves, as Schallamach himself noted:
\begin{displayquote}
    \textsl{
The waves of detachment\ldots tend to become thinner and to break up during their travel\ldots various fragments sometimes join so that individual waves gradually lose their identity. This is one of the reasons why conditions become confused in the rear part of the contact}
\end{displayquote}
This is clearly evident in the frames reproduced in Fig.~\ref{fig:orig-frames} (right), and is primarily due to the nature of the spherical indenter geometry. Despite this shortcoming, Schallamach used the spherical geometry likely because it was (and perhaps remains!) the most commonly used contact geometry by the tribology and contact mechanics community and thus has several analytical solutions \cite{Johnson_ContactMechanics, Mindlin_JApplMech_1949, Cattaneo_1938, SavkoorBriggs_1977}.

\begin{figure}[ht!]
  \centering
  \includegraphics[width=0.95\textwidth]{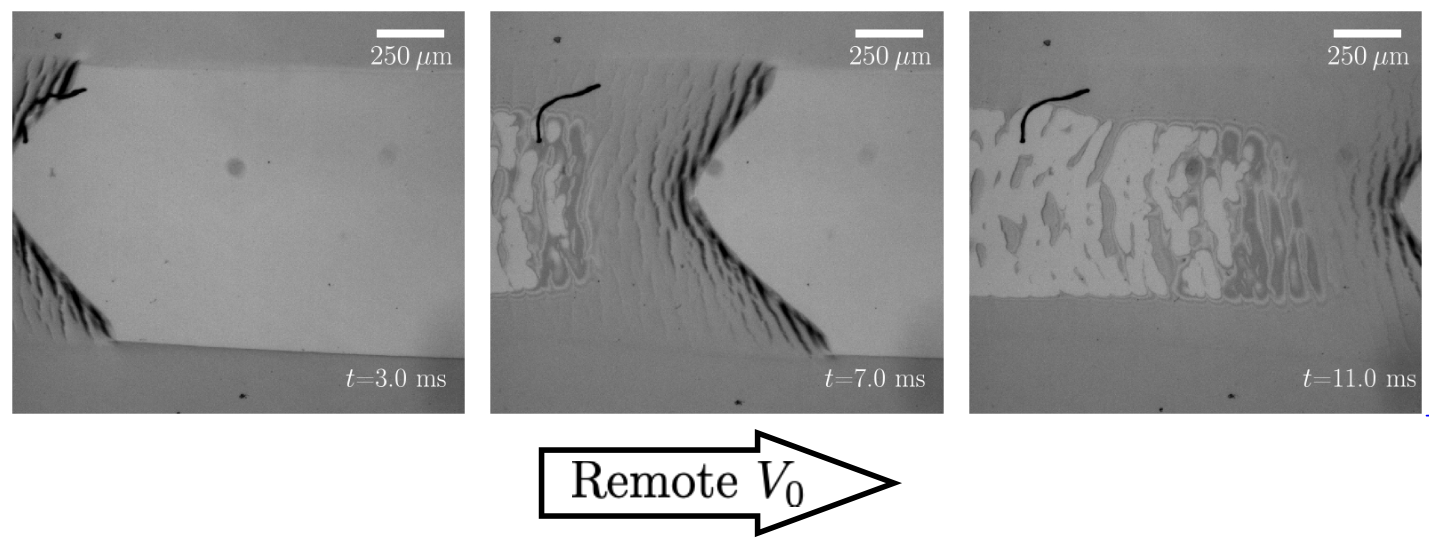}
  \caption{Evolution of a single Schallamach wave under remote applied $\sV$ to the elastic polymer. The wave propagates from left to right, in the same direction as $\sV$ and causes slip in its wake. The interface is stationary before and after the wave has passed. This is seen by following the motion of a dirt particle (black line) embedded on the polymer sub-surface. The scale bar and timestamp shows that the wavespeed is much lesser than typical elastic wave speeds.}
  \label{fig:ScW-frames}
\end{figure}

In order to isolate single waves and study their propagation dynamics, it is best to use the cylindrical indenter geometry described in Sec.~\ref{subsec:exptl}. An \lq adhesive\rq\ channel forms upon contact between the indenter and the substrate and the isolated waves can be individually imaged using high-speed videography as described earlier. Frames from a high-speed sequence showing one such wave are reproduced in Fig.~\ref{fig:ScW-frames}. Note that in this sequence, the polymer/rubber is moved from left to right at constant $V_0$ far from the interface. A single Schallamach wave traverses the interface in the same direction as $V_0$. It is clear that the defining features of the Schallamach wave remain unchanged from the spherical geometry. Firstly, the wave appears as a dark band within the interface, due to buckling-induced interface detachment and readhesion. Secondly, points on the interface remain stationary until the wave appears; they then begin to move and are displaced by a fixed amount $\Delta x$. Finally, the wave is accompanied by compression wrinkles, just as we observed in the case of the spherical contact geometry, \emph{cf.} Fig.~\ref{fig:wrinkles}.

%The images of this solitary Schallamach wave allow us to determine its propagation speed more quantitatively via the use of space--time diagrams \cite{ViswanathanETAL_SoftMatter_2016_1}. . 

In light of these single wave results, it is pertinent to revisit Schallamach's observations about \lq imperfections\rq\ in the spherical contact itself:
\begin{displayquote}
\textsl{
\ldots secondary effects occur at times\ldots where a relatively large portion of the contact becomes detached from the main body and suddenly fuses with it again.}
\end{displayquote}
This statement refers to the rear end of the contact (taken in the sliding direction). Just as material at the front end of the contact is subject to compression and surface buckling, the surface just outside this rear end is subject to tension. One can consequently imagine a local \lq tensile neck\rq\ causing detachment from the indenter surface, followed by readhesion. These are the \lq secondary effects\rq\ that Schallamach refers to. They have since been backed up by similar observations in spherical contacts of various dimensions \cite{Barquins_Wear_1985, Barquins_Wear_1993}.

The equivalent of this seemingly innocuous observation in our cylindrical contact geometry of Fig.~\ref{fig:schematic-adhesive} is dramatic. If a tensile neck does indeed form just outside the contact zone, resulting in a locally detached zone within the interface, it must propagate in the opposite direction in the form of another, very different, wave of detachment!

\begin{figure}[ht!]
  \centering
  \includegraphics[width=0.95\textwidth]{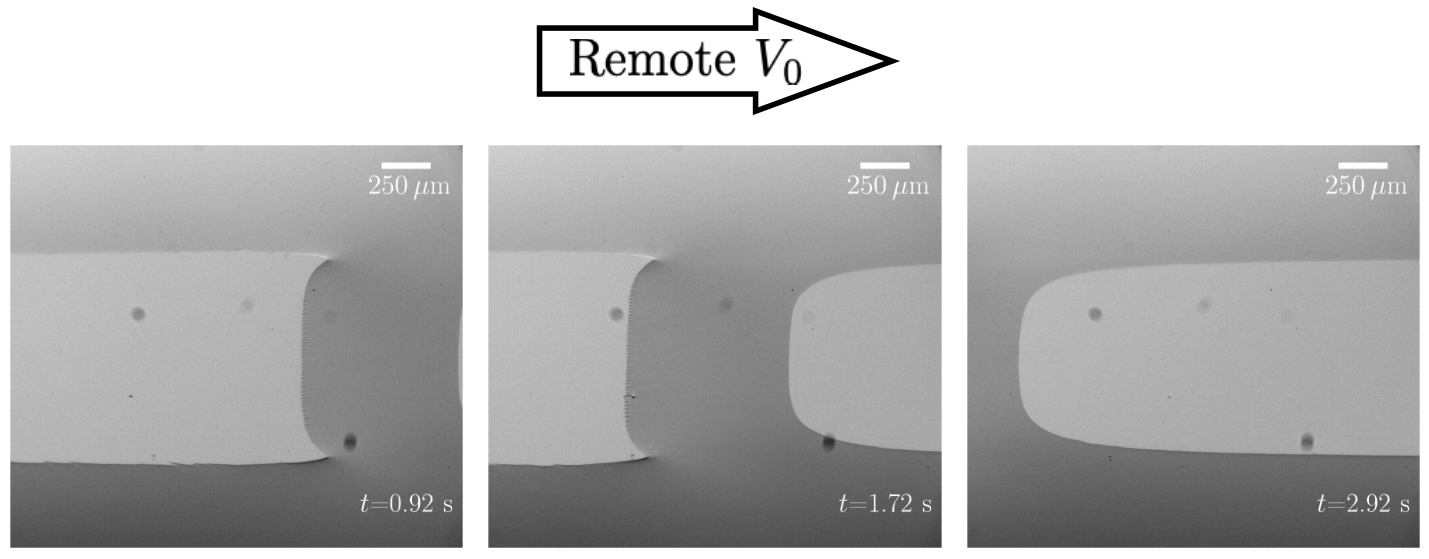}
  \caption{Propagation of the dual of the Schallamach wave or the separation pulse. Remote sliding $\sV$ occurs in the same direction as before, but the detachment wave moves in the opposite direction. Despite this, slip occurs in the same direction as the applied $\sV$. Note that, just as with the Schallamach wave, this separation pulse also propagates at speeds much lesser than elastic wave speeds.}
  \label{fig:SP-frames}
\end{figure}

One such instance is illustrated by the set of frames taken from a high--speed sequence shown in Fig.~\ref{fig:SP-frames}. Here again, the remote sliding direction $V_0$ is from left to right, just as in Fig.~\ref{fig:ScW-frames}. However, a detachment wave is now seen to move in the opposite direction (right to left), yet causing slip in the same direction as before. The latter fact may be ascertained by the fact that a fortuituously trapped dirt particle within the contact is translated in the direction of remote $V_0$, as is to be expected. Furthermore, this wave of detachment is easily distinguished from the Schallamach wave by a complete absence of compression-induced surface wrinkles---it is nucleated by a tensile instability after all---resulting in perfect contact in its wake. This detachment wave has been christened the separation or detachment pulse \cite{ViswanathanETAL_SoftMatter_2016_1}. For the purposes of this article, we also refer to this as the dual of the Schallamach wave, since it differs only in that it is tensile and propagates in the opposite direction to the latter.

The compressive \emph{versus} tensile nature of the Schallamach wave and its dual, the separation pulse, is sufficient to explain the corresponding propagation direction observed in Figs.~\ref{fig:ScW-frames},~\ref{fig:SP-frames}, see Ref.~\cite{ViswanathanETAL_SoftMatter_2016_2}. For wave propagation at constant speed $c_w$ along the $x-$axis, the tangential displacement $u_x$ has the form $u_x(x - c_w t)$. Hence, the $x-$derivative of the displacement $\partial u_x/\partial x$ and the strain $\epsilon_{xx}$ can be written as
\begin{equation}
  \label{eqn:propDir}
	\epsilon_{xx} = \frac{\partial u_x}{\partial x} = -c_w^{-1} \frac{\partial u_x}{\partial t} \simeq -c_w^{-1} \frac{\Delta x}{T}
\end{equation}
where $\Delta x$ is the slip per wave and $T$ is duration of propagation. Since the sign of $\Delta x$ is the same as $\sV$, the strain $\epsilon_{xx}$ is negative (compressive) if the wave propagates in the same direction as $\sV$ and is positive (tensile) if it propagates in the opposite direction. Therefore, for the case of Schallamach waves $\epsilon_{xx} <0$ and for separation pulses $\epsilon_{xx} > 0$, as is evident from the frames in Figs.~\ref{fig:ScW-frames} and~\ref{fig:SP-frames}.

\subsection{Interface dynamics from space-time diagrams}
\label{subsec:spaceTime}

\begin{figure}[ht!]
  \centering
  \begin{minipage}{0.3\textwidth}
  \includegraphics[width=\textwidth]{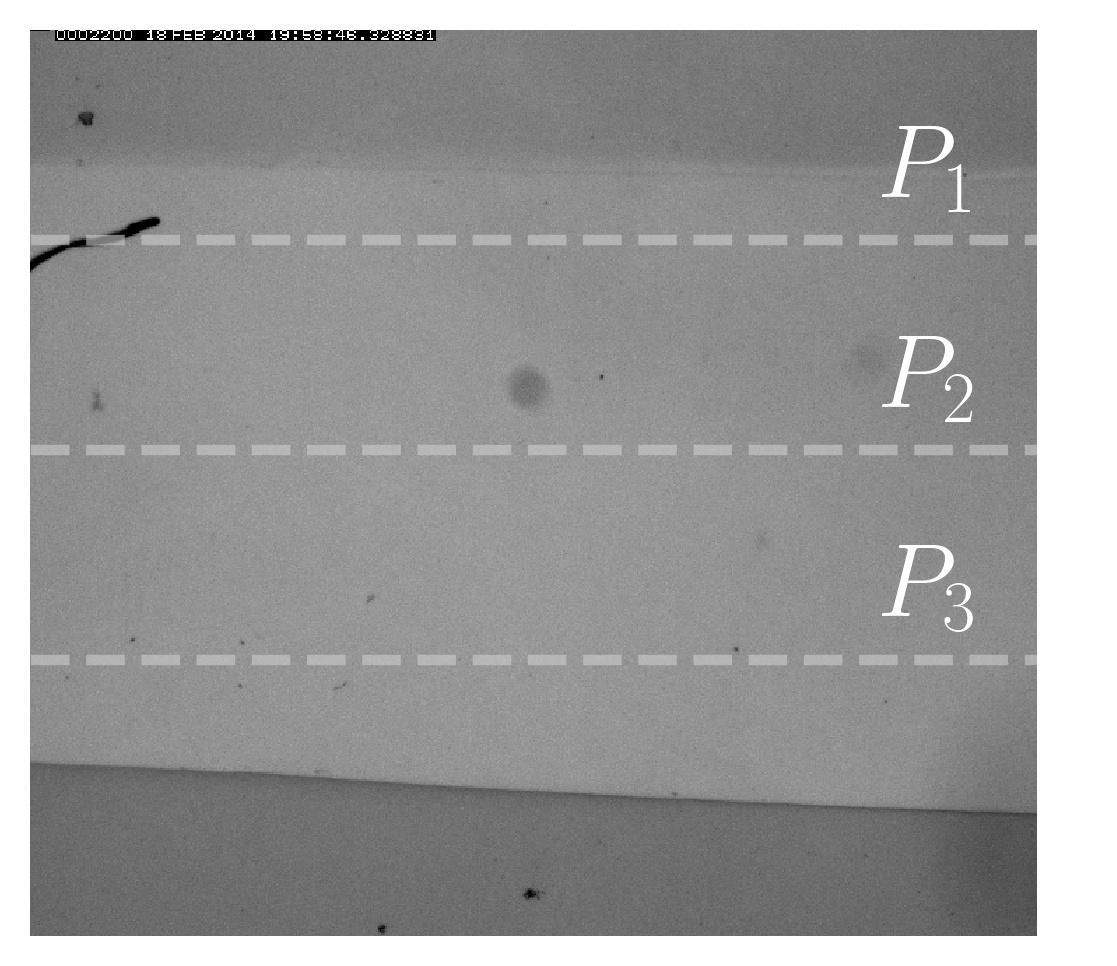}  
  \end{minipage}
  \begin{minipage}{0.65\textwidth}
    \includegraphics[width=\textwidth]{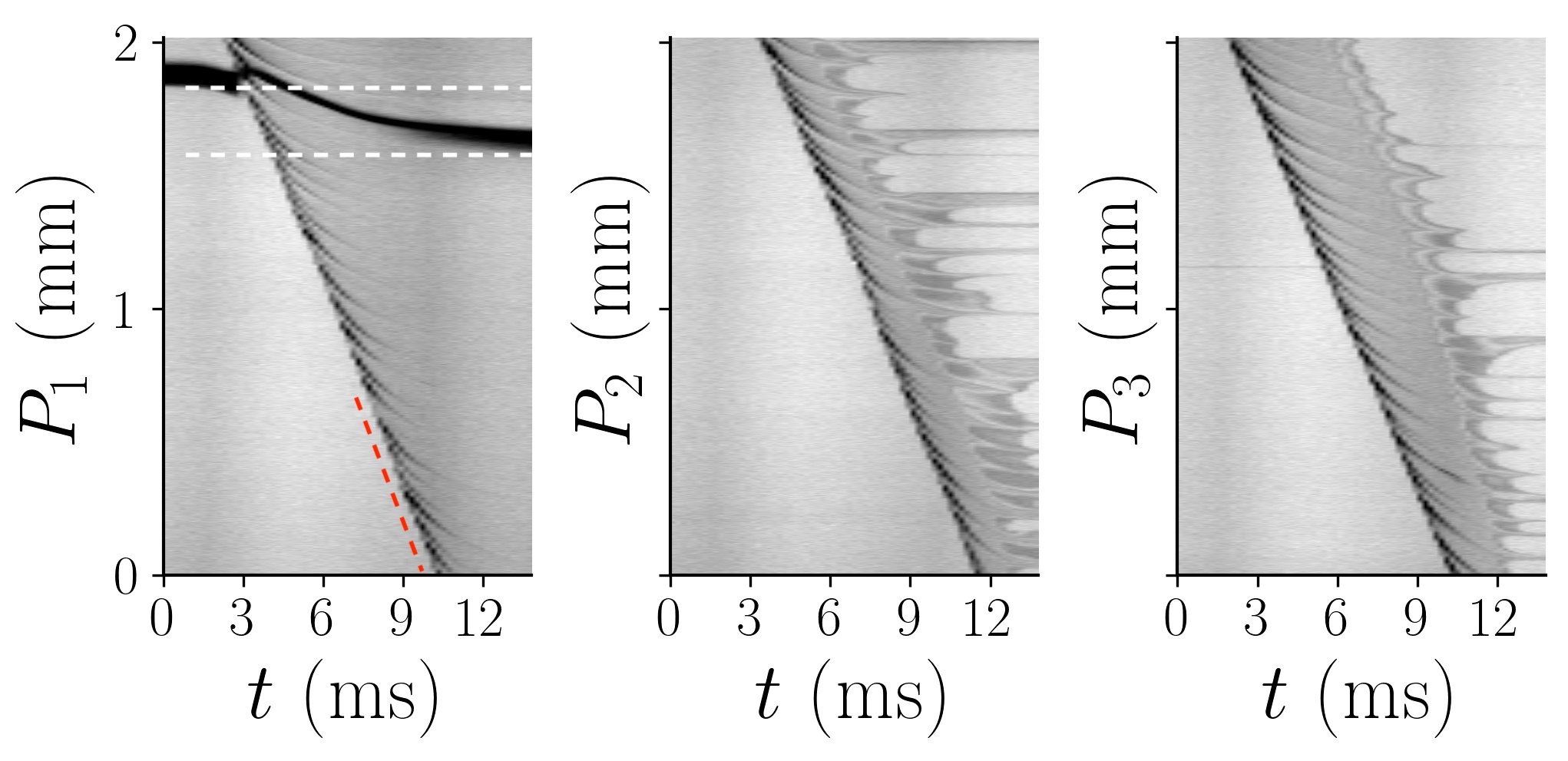}
  \end{minipage}
  \quad
  \begin{minipage}{0.3\textwidth}
  \includegraphics[width=\textwidth]{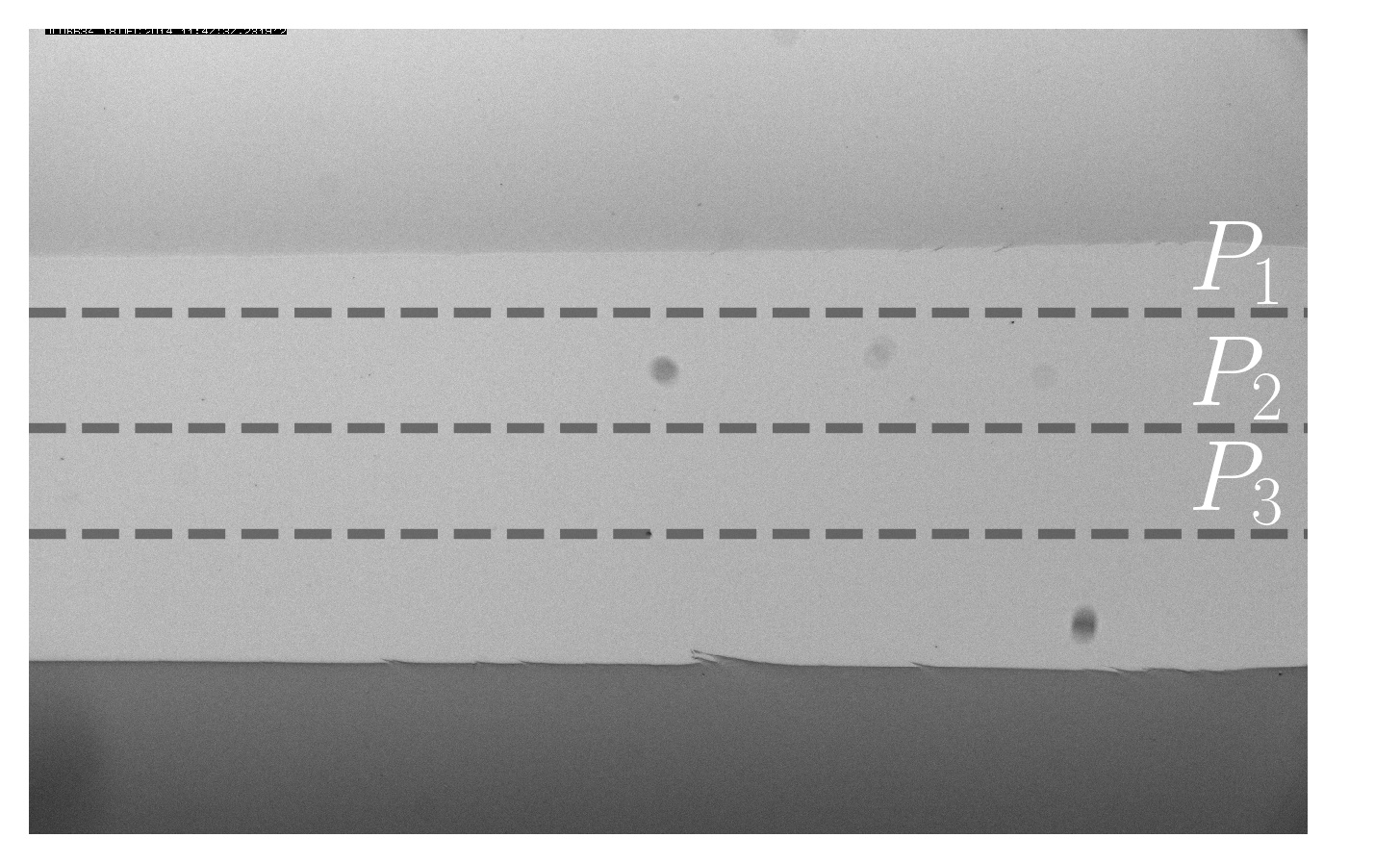}  
  \end{minipage}
  \begin{minipage}{0.65\textwidth}
    \includegraphics[width=\textwidth]{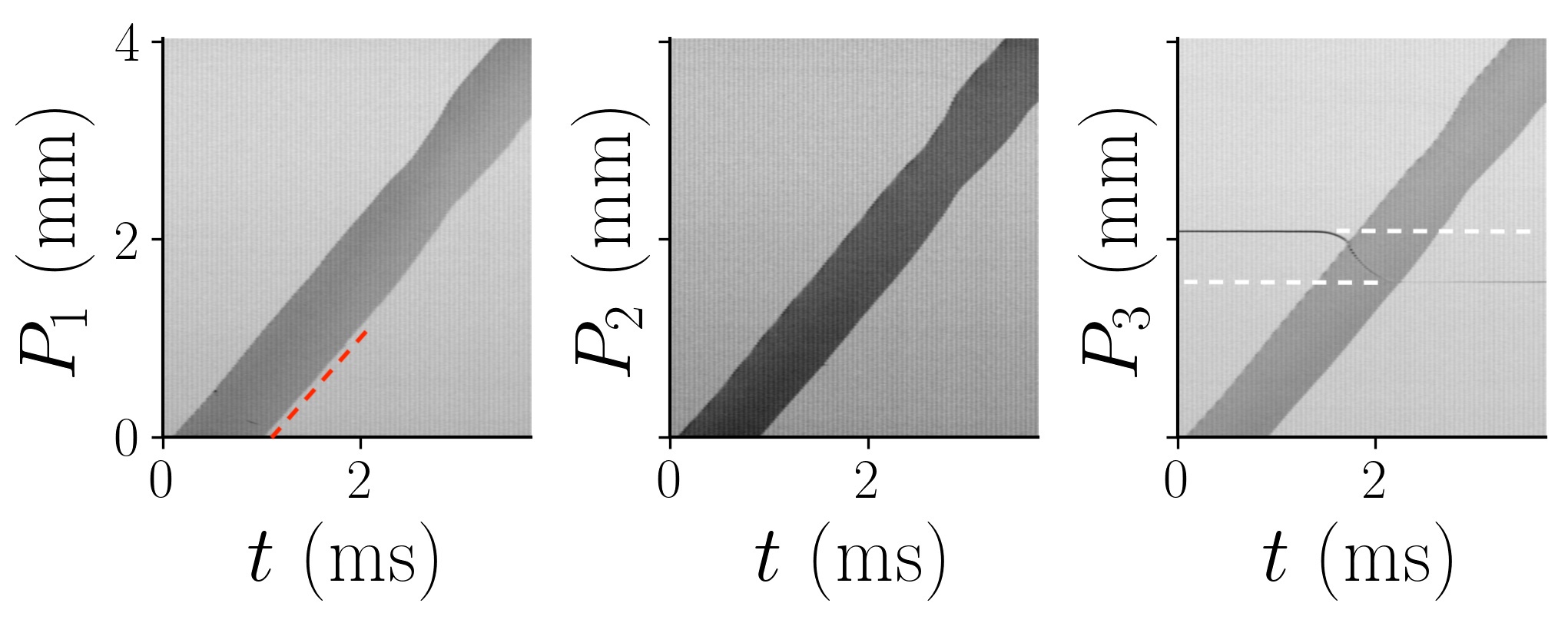}
  \end{minipage} 
  \caption{Space--time diagrams showing propagation of a Schallamach wave (top row) and its dual (separation pulse, bottom row). The image on the left shows locations of the horizontal lines $P_1, P_2, P_3$, with corresponding space--time diagrams to the right in each row. The wave appears as a dark band in all three diagrams with constant slope (red dashed line) equal to the wave speed. Interface slip distance, as indicated by the trajectory of the black line marker, is as shown between parallel white dashed lines. Analogously, the separation pulse appears as a dark band in all three diagrams (bottom row) with constant slope (red dashed line) equal to the wave speed. Interface slip distance, as indicated by a moving dirt particle near the interface, is marked by parallel white dashed lines.}
  \label{fig:spaceTime}
\end{figure}

\mt{More quantitative information about interface dynamics accompanying wave motion is obtained by means of space-time diagrams, that are commonly used in wave propagation studies to obtain spatio-temporal information \cite{Whitham_1975}.} A space-time diagram is constructed by stacking a fixed horizontal line in the image sequence as a function of time. Three such fixed lines, denoted $P_1, P_2, P_3$, and their corresponding space-time diagrams are presented in Fig.~\ref{fig:spaceTime} for Schallamach (top row) and separation waves (bottom row). In each case, the wave itself appears as a dark diagonal band with a constant inclination angle (red dashed line). Based on these diagrams, we can assert that both the Schallamach wave and its dual propagate at constant speed through the interface. The former is in the range $c_w \sim 100-300$ mm/s, corresponding to a sliding velocity $V_0 = 50-250\,\mu$m/s and the latter is in the range $\sim 10$ mm/s with $\sV =10-50\,\mu$m/s. Clearly, $c_w \gg V_0$ in both cases, yet is much lesser than any elastic wave speed for the elastic material ($\sim 100$ m/s).

The differences between the two waves is easily seen in the sign of the slope in these diagrams. The negative (positive) slope is due to the corresponding wave moving in the same (opposite) direction as applied $\sV$. Additionally, the interface slip $\Delta x$ is also self-evident: for Schallamach waves, this can be obtained by tracking the dark patch (top panel) in frame $P_1$ of the space-time diagram. The patch has slipped by a constant finite distance (between white dashed lines) due to wave motion. On the other hand, a dirt particle just below the surface may be tracked for the separation pulse (frame $P_3$, bottom row) which shows the resulting slip due to wave passage (again between white dashed lines).

\subsection{Wave motion leading to stick--slip friction}

\begin{figure}
  \centering
  \includegraphics[width=0.45\textwidth]{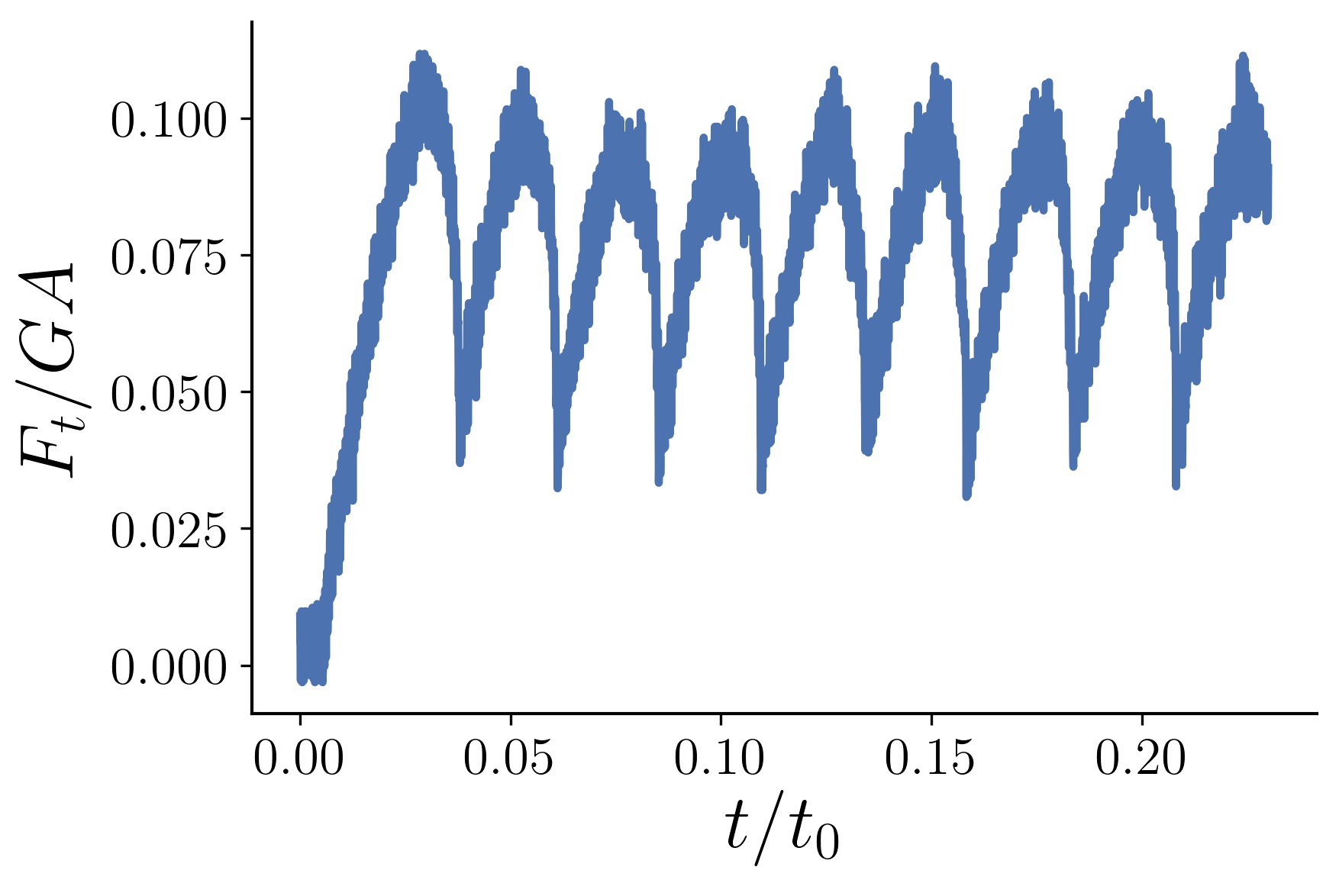}
  \quad\quad
  \includegraphics[width=0.45\textwidth]{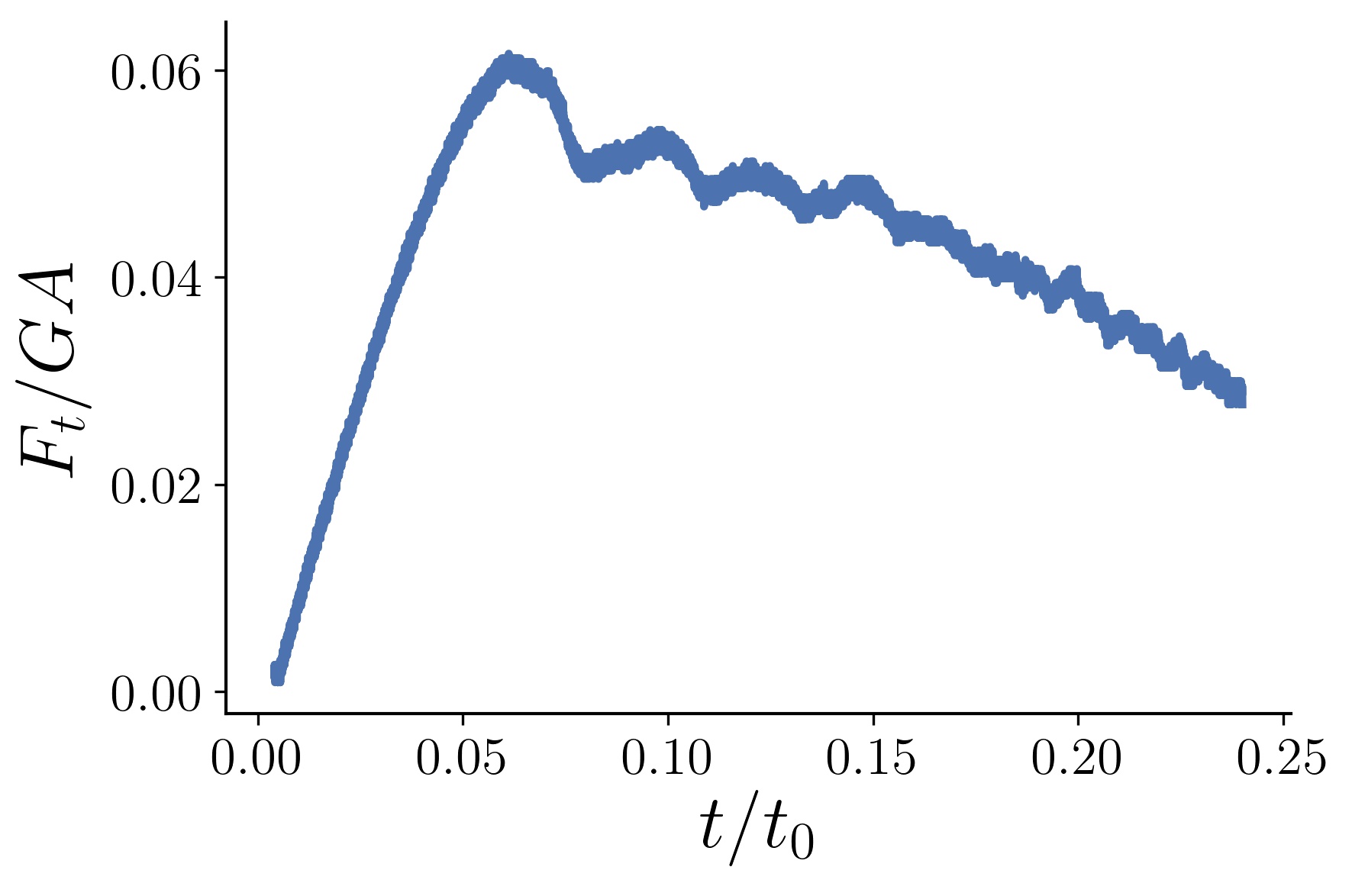}
  \caption{Forces accompanying the propagation of a sequence of Schallamach waves (left) and separation pulses (right). Both traces show oscillations, and are therefore characteristic of what might be termed \lq stick--slip\rq\ motion. Wave nucleation occurs at the peak in each oscillation cycle, followed by a force reduction during propagation. The magnitude of the reduction, as well as the wave nucleation frequency differs between the Schallamach wave (left) and its dual (right). Note that force is non-dimensionalized by the product of shear modulus $G$ times contact area $A$ and time is non-dimensionalized by $t_0 = L/\sV$ where $L$ is the contact length and $\sV$ is the sliding velocity as before.}
  \label{fig:ScW-SP-forces}
\end{figure}

Since the experiments described so far have been performed with a constant remote velocity boundary condition, it is instructive to look at the corresponding time-dependent tangential force. Forces for the Schallamach wave and the separation pulse are shown in Fig.~\ref{fig:ScW-SP-forces}(a) and (b), respectively. Time on the horizontal axis is non-dimensionalized by a characteristic sliding time $t_0 = L/V_0$, with $L$ being the length of the contact. Force on the vertical axis is non-dimensionalized by the product of shear modulus $G$ and nominal contact area $A$. It is clear at once that both force traces can be termed as characterizing \lq stick--slip\rq\ motion at the interface. At the start of sliding, both traces increase from zero to a maximum value at which wave nucleation occurs. Ensuing wave propagation triggers a sharp reduction in the tangential force. Once a single wave has traversed the length of the contact, the process repeats; the attendant interface motion is thus intermittent, mediated only by the propagation of single Schallamach waves or separation pulses. On the contrary, the two traces have very different quantitative features, such as the frequency and reduction magnitude per cycle, see Fig.~\ref{fig:ScW-SP-forces}.

It is indeed surprising that the tensile dual of the Schallamach wave had not been reported for a very long time, even though signs of their nucleation were noted very early on by both Schallamach and Barquins. In fact, tensile separation pulses have even been incorrectly reported as Schallamach waves in the geophysics community \cite{YamaguchiETAL_JGeoPhysRes_2011}. The likely cause for this is the widespread use of the spherical indenter geometry which obscures observations of single/solitary wave motion.

\begin{figure}[ht!]
  \centering
  \includegraphics[width=0.7\textwidth]{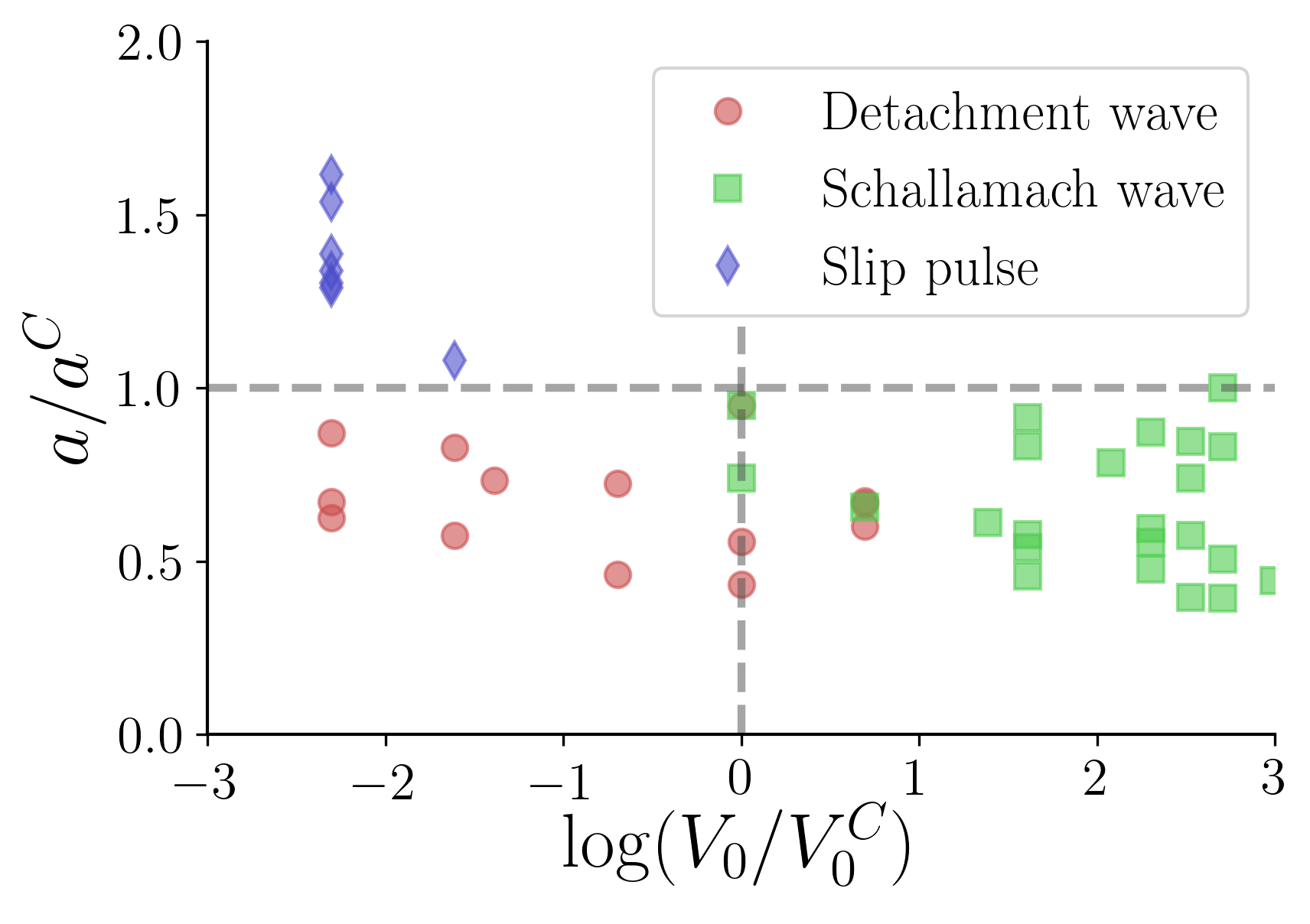}
  \caption{A two-parameter phase diagram showing domains of occurrence of the three Schallamach-type waves discussed in the article. Schallamach waves occur above a critical sliding velocity, separation pulses occur a low $\sV$ and normal loads while detachment-less slip pulses mediate motion at large normal loads.}
  \label{fig:phaseDig_Expt}
\end{figure}

Together, these two waves are mechanically the only possibilities for moving detachment waves at soft elastic interfaces characterized by adhesion. If the need for physical detachment of the interface is relaxed, one other wave is also commonly observed---the so-called slip pulse \cite{BaumbergerETAL_PhysRevLett_2002, Caroli_PhysRevE_2000}. This peculiar wave is also nucleated in compression so that by Eq.~\ref{eqn:propDir}, it travels in the same direction as a Schallamach wave but looks nothing like it. Given the absence of any detachment, the existence of the slip pulse can only be ascertained using photoelastic techniques to directly measure the interface shear stress \cite{ViswanathanETAL_SoftMatter_2016_2, ViswanathanSundaram_Wear_2017}.

Taken together, these three waves---the Schallamach wave, its dual the separation pulse and the slip pulse---appear to be the sole mediators of stick--slip motion in low velocity sliding. The domains of occurrence of each of these waves in the $\sV-a$ plane is shown in Fig.~\ref{fig:phaseDig_Expt}. The control parameters in this \lq phase diagram\rq\ are $\sV$ and the normal load (equivalently contact width $a$). At low $\sV$ and $a$, stick--slip occurs via the propagation of separation pulses, while beyond a certain velocity threshold $V_0^C$ (vertical dashed line), Schallamach waves predominate. The likely reason for this is the ease with which the corresponding nucleation events can occur. On the contrary, with larger normal loads/contact widths above a threshold $a^C$, both tensile detachment and compressive buckling become infeasbile so that the slip pulse is the primary cause for stick--slip. It now appears fairly certain that these three waves are the sole mediators of low velocity stick--slip in deformable adhesive interfaces.

\subsection{Similarities with elastic dislocations: Burgers' vector and defect pinning}
\label{subsec:dislocations}

At this junction, it is instructive to take a step back and reconsider the basic defining feature of a Schallamach-type wave, \emph{viz.} that it propagates through the interface at a constant speed, and causes slip in its wake. Based on this minimal description alone, one would be hard-pressed to exclude crystal lattice dislocations from this set as well. Dislocations are also similarly nucleated, propagate at finite speed, and cause (plastic) slip in their wake. 
\begin{figure}
  \centering
  \includegraphics[width=0.6\textwidth]{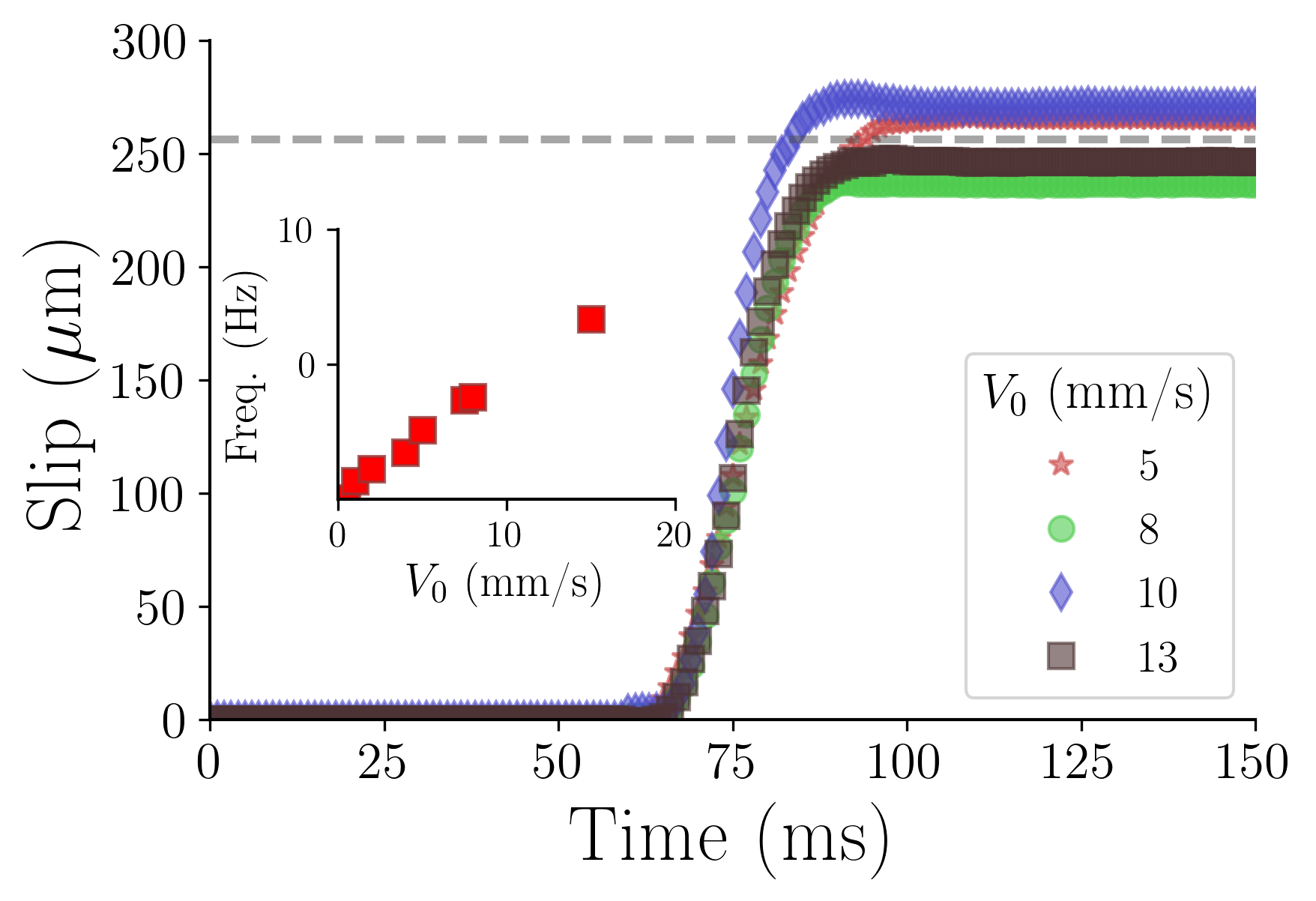}
  \caption{Analogies between solitary Schallamach-type waves and crystal dislocations. Time-dependent displacement curve as recorded at a single point location on the interface. This curve shows that the point remains stationary until a wave approaches $(\sim 13$ ms), following which slip occurs and reaches a maximum value $\sim 250\,\mu$m. This is identical with what happens with a gliding crystal dislocation so that the effective slip may be identified with the magnitude of the Burgers' vector. Inset shows the frequency of wave generation plotted against $\sV$, the proportionality is again reminiscent of the plastic strain rate relation due to dislocation motion.}
  \label{fig:dislocation}
\end{figure}

For concreteness, consider the solitary Schallamach wave introduced in Sec.~\ref{subsec:solitarySchallamach}. Three distinct velocity scales are evident---the imposed (remote) sliding velocity $\sV$, the local material velocity $\vec{v}_p$ due to the wave and the velocity $c_w$ of propagation of the wave itself. Directions of $\sV$ and $c_w$ are fixed by the external loading. While $c_w$ is constant as the wave propagates through the interface (see Fig.~\ref{fig:spaceTime}), the local material velocity $\vec{v}_p$ varies with time from zero (as the wave approaches), to a maximum value and finally back to zero (as the wave departs). For example, for the data in Fig.~\ref{fig:spaceTime}, instantaneous $\vec{v}_p$ is determined by the slope of the dirt particle curve at that time; $c_w$ is the slope of the band (constant) and both are indeed distinct from $\sV$.

Based on data such as that in Fig.~\ref{fig:spaceTime}, the relative displacement of material points on the surface can be plotted as a function of time, see Fig.~\ref{fig:dislocation}. This plot shows the displacement of a single point under different remote $\sV$ due to the propagation of a single Schallamach wave at some corresponding speed $c_w$. Three features are immediately obvious from this data: firstly, that the surface is displaced in the form of a unit step displacement $\vec{b}$. We had earlier denoted this slip distance as $\Delta x = |\vec{b}|$ that occurs in the direction of $\sV$. Secondly, $|\vec{b}|$ is independent of both $\sV$ and $c_w$ since curves for a range of values result in similar surface slip. This is understandable within the buckling framework discussed in Sec.~\ref{sec:exptl}---the buckling threshold is independent of $\sV$. Finally, since the interface moves only via such slip events, the remote $\sV$ has to be accommodated by waves with a frequency $n$ such that:
\begin{equation}
  \label{eqn:strainRate}
  \sV = n |\vec{b}|
\end{equation}
The constancy of $|\vec{b}|$ in this equation implies that $n \propto \sV$. This is indeed the case, as supported by the data presented in the inset graph to Fig.~\ref{fig:dislocation}. With $\sV$ ranging from $10^{-1} - 20$ mm/s, a linear dependence between the wave generation frequency $n$ and $\sV$ is clearly established.

The situation summarized in Fig.~\ref{fig:dislocation} also applies equally well to the motion of dislocations (via slip) on a crystal plane \cite{Nabarro_CrystalDislocations_1967}. In this case, the displacement magnitude is given by the dislocation Burgers vector (hence the notation $\vec{b}$ above) and corresponds to the jump in Fig.~\ref{fig:dislocation}. One can hence define this as the \lq Burgers' vector\rq\ of a Schallamach wave. It is determined, for a given contact geometry, by the substrate material properties---for the experimental configuration in Sec.~\ref{sec:exptl}, the mean $|\vec{b}| =255$ $\mu$m.  Additionally, Eq.~\ref{eqn:strainRate} appears analogous to the Taylor-Orowan equation for plastic strain rate due to dislocation motion.

\begin{figure}
  \centering
  \includegraphics[width=0.5\textwidth]{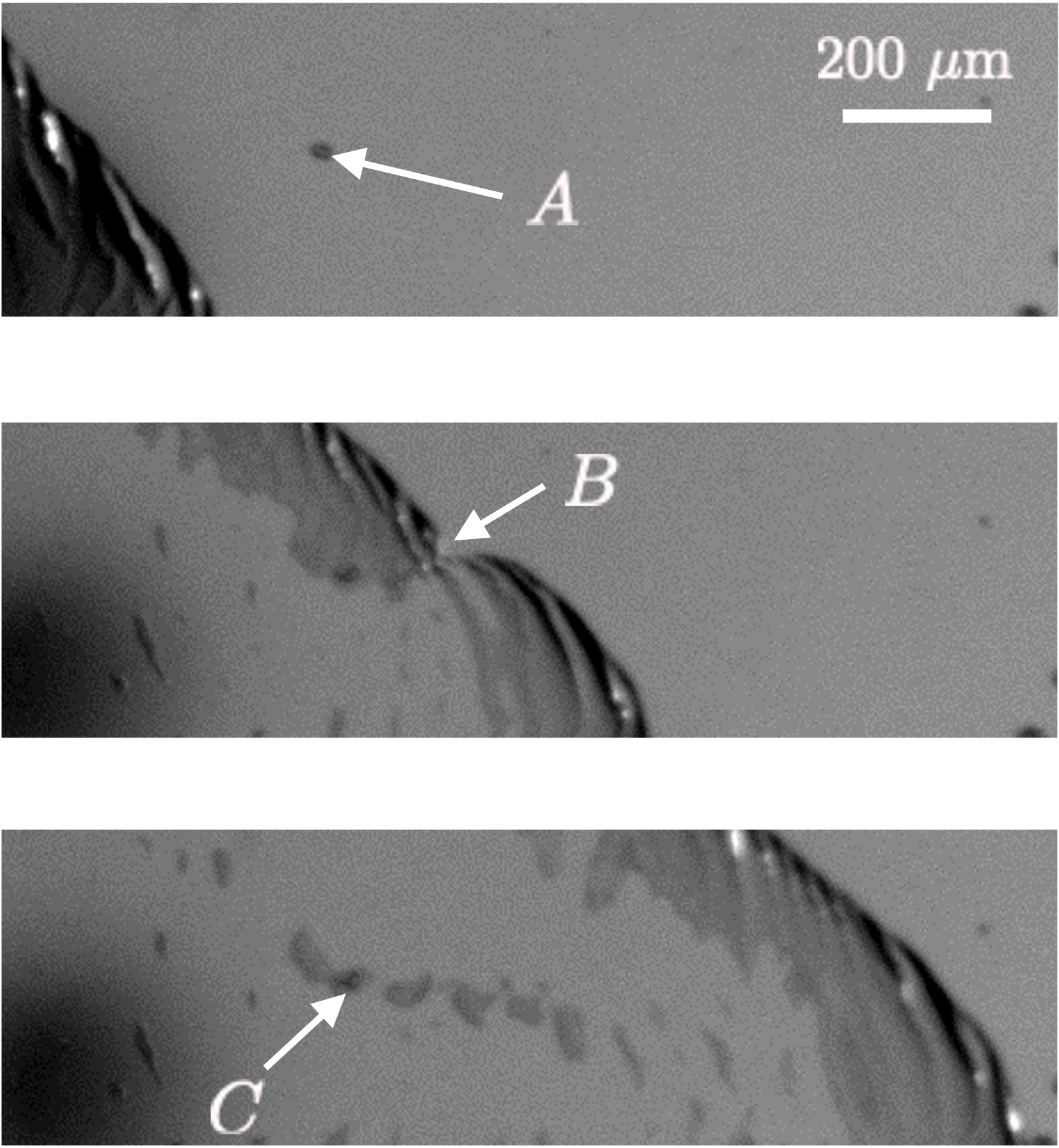}
  \caption{Interaction between a solitary Schallamach wave and a large dirt particle $A$ on the interface. The wave approaches the particle and is bowed by it (at $B$), following which a residual detachment region is left behind $(C)$ around the dirt particle. This outcome is also analogous to that observed during dislocation-point defect interaction.}
  \label{fig:dirt}
\end{figure}

The analogy with dislocations is further strenghtened by an interesting phenomenon observed during propagation of Schallamach waves past stationary dirt particles attached to the indenter, see Fig.~\ref{fig:dirt}. Here, the wavefront approaches a static dirt particle $A$ in the contact region and gets \lq pinned\rq , causing a bend in the wave profile (point $B$). As the front moves away from the particle, the wave regains its original profile, leaving a residual air pocket $C$ around the particle. This \lq bowing\rq\ effect also occurs when dislocations move past static obstacles (e.g., solute particles), leaving behind Orowan loops \cite{Nabarro_CrystalDislocations_1967}. Under dilute solute particle concentration, the bent dislocation line eventually relaxes and regains its shape.

%For dislocations in metals, this is known to lead to the Fisher, Hart and Pry (FHP) effect \cite{Cottrell_MechPropMatter}.

%The observed propagation characteristics of solitary Schallamach waves help expand on the analogy between wave propagation and crystal dislocation glide. Firstly, Schallamach waves are nucleated at a critical stress, see Fig.~\ref{fig:ScW-SP-forces}, similar to crystal dislocations. This stress is the compression required for buckling to occur on the elastomer free surface. Secondly, slip at the interface determines an equivalent Burgers vector $\mathbf{b}$ for the Schallamach wave (\emph{cf.} Fig.~\ref{fig:dislocation}(left)) with $|\mathbf{b}|$ independent of $V_0$. Thirdly, the frequency with which Schallamach waves are generated is analogous to the strain rate resulting from repeated dislocation motion at the interface, sample supporting measurements are presented in Fig.~\ref{fig:dislocation}(right). This is quite reminiscent of the Orowan equation relating strain rate to Burgers vector for dislocation glide \cite{Nabarro_CrystalDislocations_1967}.   

\section{How can we describe Schallamach-type waves?}
\label{sec:theory}

We now turn our attention to continuum descriptions of Schallamach-type waves that occur at the interface between two contacting solids. Unlike in fluids, waves at solid surfaces/interfaces can occur in various forms/guises depending on the nature of the boundary conditions \cite{Achenbach_WavePropagation_2012}. A primer for the propagation of surface waves in elastic media is presented in the Appendix. In this section, we explore the predictions of an elastodynamic formulation for Schallamach-type wave propagation.

\subsection{Mixed-boundary value problem for Schallamach-type waves}
\label{subsec:continuum}

In addition to the simple boundary conditions discussed in the appendix, interface waves often propagate under more complex, often mixed, boundary conditions. The occurrence of net interface slip due to a wave further complicates this situation. Slip-less detachment waves were analyzed by Comninou and Dundurs, and found to propagate with a range of velocities, all comparable to the Rayleigh wave speed \cite{ComninouDundurs_JApplMech_1977, ComninouDundurs_JApplMech_1978}. On the other hand, detachment-free frictional waves have also been analyzed using Coulomb-friction boundary conditions. However, these are marred by inconsistencies in enforcing the static to dynamic transition, so that a physically valid solution does not presently exist for any combination of materials \cite{MartinsETAL_JVibrAcous_1995, Adams_JApplMech_1998, RanjithRice_JMechPhysSolids_2001}. In addition, the biggest common shortcoming of all these solutions is that they cannot account for interface slip in the wake of the wave. 

It is thus clear that none of these solutions can describe the primary features of Schallamach-type waves---slow propagation velocity and net interface slip---with or without interface detachment. The first feature, that the wave velocity $c_w \ll c_T, c_L$, suggests the existence of another velocity-scale, likely due to a velocity-dependent friction law \cite{Rabinowicz_SciAm_1956, Dieterich_JGeophysRes_1979, Ruina_JGeophysRes_1983, li2011frictional} and/or some type of regularization \cite{CochardRice_JGeophysRes_2000, BaumbergerCaroli_AdvPhys_2006}. Yet, introducing a velocity scale via a suitable interface friction law is non-trivial. \mt{While this is possible with rather simplistic 1D models \cite{BrenerETAL_EurPhysJE_2005}, a general description for moving detachment waves with adhesion has hitherto remained elusive.}

A recently established elastodynamic framework for describing Schallamach-type detachment waves has, however, shown significant promise \cite{ViswanathanETAL_SoftMatter_2016_2}. \mt{We present a 2D version of the quasi-3D framework presented in this work, the sequence of algebraic steps described below may be reconstructed by taking recourse to Ref.~\cite{ViswanathanETAL_SoftMatter_2016_2}. Consider the elastic material occupying the half-space $z\leq 0$. In order to mimic the physical conditions accompanying wave propagation, this framework uses a mixed boundary condition formulation involving stationary and detached zones on the contact interface $z=0$. For a wave moving along the positive $x$-axis with velocity $c_w$, one can use co-moving coordinates $\eta = k(x - c_w t)$ to make the wave appear stationary. The wave number $k$ is related to both the wave width 2$\alpha$ and the parameters $V_0$ (sliding speed), $c_w$ \cite{ViswanathanETAL_SoftMatter_2016_2}} Assuming deformation to be restricted to the $xz$-plane, the relevant mixed boundary conditions are:
\begin{align}
  \label{eqn:BCs}
  \text{Detachment zone } |\eta| &< \alpha
  \begin{cases}
    &\sigma_{zz} = \sigma_{zx} = 0\\
    & \dot{u}_x(\eta), \dot{u}_z(\eta) \neq 0
  \end{cases}\\
  \text{Stationary/stick zone } \alpha < |\eta| &< \pi
  \begin{cases}
    &\sigma_{zz}(\eta), \sigma_{zx}(\eta) \neq 0\\
    & \dot{u}_x = \dot{u}_z = 0
  \end{cases}
\end{align}
where the $\sigma_{xz}, \sigma_{zz}$ and $\dot{u}_x, \dot{u}_z$ are the interface stresses and velocities, respectively. \mt{The unknowns here are the stresses $\sigma_{xx}, \sigma_{xz}$ in the sticking zone and the velocities $\dot{u}_x, \dot{u}_z$ in the detached zone. In order to obtain expressions for these unknowns, we begin with a commonly used \emph{ansatz} for surface wave solutions in the form of a dual series representation, see for e.g., Ref.~\cite{ComninouDundurs_JApplMech_1977}. The boundary conditions in Eq.~\ref{eqn:BCs} are then used to obtain coupled singular integral equations for the interface velocities $\dot{u}_x, \dot{u}_z$ using the procedure discussed in Ref.~\cite{ViswanathanETAL_SoftMatter_2016_2}. In order to address the slow velocity scale issue, one exploits the fact that $c_w \ll c_T, c_L$ (elastic wave speeds) to perform a perturbation expansion in powers of $c_w/c_L$ and $c_w/c_T$. The result is that both Schallamach waves and the oppositely moving separation pulse can be described within the same elastic framework. In the limit $\nu \to 0.5$ (as with incompressible rubbers), $\dot{u}_x(\eta), \dot{u}_z(\eta)$ can be shown to obey the following singular integral equations:
  \begin{align}
    \label{eqn:SIEs}
    \frac{\tau_r}{G} + \frac{2}{\pi} \bigg[\text{sec}^2(\eta/2)\int_{-\alpha}^\alpha \dot{u}_x(\eta^\prime) \cot\left(\frac{\eta - \eta^\prime}{2}\right) d\eta^\prime  + \frac{\pi \tan(\eta/2) V_0}{c_w}\bigg]  &= 0\\
    \frac{\sigma_r}{G} + \frac{2}{\pi} \bigg[\text{sec}^2(\eta/2)\int_{-\alpha}^\alpha \dot{u}_z(\eta^\prime) \cot\left(\frac{\eta - \eta^\prime}{2}\right) d\eta^\prime\bigg]  &= 0
  \end{align}
  where $\sigma_r,\tau_r$ are the remote normal and shear stresses, $G$ is the material's shear modulus and $\eta = k(x-c_w t)$ as described earlier.}
  
\mt{Unlike most other analytical treatments of stick--slip waves, the boundary conditions Eq.~\ref{eqn:BCs} and final integral equations (Eq.~\ref{eqn:SIEs}) imply that a description of Schallamach-type waves does not need either an explicit friction law (e.g., Coulomb or rate/state) or a velocity scale.} Further, since the boundary conditions in Eq.~\ref{eqn:BCs} are written in terms of velocities, interface slip can be enforced by suitably setting the integration constant in the expressions for the final displacements.

\mt{The unknowns in Eq.~\ref{eqn:SIEs} can be obtained by inverting the singular integral equations, resulting in expressions for the interface velocities and, upon integration, the displacements \cite{ViswanathanETAL_SoftMatter_2016_2}. Of particular interest are the interface stresses, which can also be expressed in closed form using expressions for the displacements and Hooke's law
  \begin{align}
    \label{eqn:stresses_1}
    \frac{\sigma}{\sigma_r} &= \frac{|\tan(\eta/2)|}{\sin(\alpha/2)} \frac{1}{\sqrt{\tan^2(\eta/2) - 1}} \\\label{eqn:stresses_2}
    \frac{\tau}{\tau_r} = \frac{|\tan(\eta/2)|}{\sin(\alpha/2)} \frac{1}{\sqrt{\tan^2(\eta/2) - 1}} &- \left(\frac{V_0}{c_w\tan(\alpha/2)}\right)\left(\frac{2G}{\tau_r}\right) \frac{|\tan(\eta/2)|}{\tan(\eta/2)} \frac{1}{\sqrt{\tan^2(\eta/2) - 1}}
  \end{align}
  and apply to both the Schallamach wave and its dual, by suitable change in the sign of the ratio $c_w/V_0$. }

The three main predictions of this theoretical treatment are as follows: \mt{Firstly, Eqs.~\ref{eqn:stresses_1} and~\ref{eqn:stresses_2} determine interface stresses accompanying both the Schallamach wave and its dual. In fact, Eq.~\ref{eqn:stresses_1} shows a singularity near the detachment zone edge $\eta \to \pm \alpha$, implying that some adhesion mechanism is operative here \cite{JohnsonETAL_ProcRoySocA_1971}. This is noteworthy since the entire problem, from the boundary conditions to the final integral equations did not specity an adhesion law \emph{a priori}. Adhesion is hence a necessary condition for such wave propagation. Secondly, wave motion can occur over a range of speeds $c_w/V_0$ as opposed to a single, characteristic speed. This shows that slow-moving Schallamach-type waves can be described without having to introduce an explicit velocity scale.} 

\mt{Finally, both stress components in Eq.~\ref{eqn:stresses_1},~\ref{eqn:stresses_2} show square-root singular behaviour near the detachment zone edges $\eta \to \pm \alpha$. The normal stress component yields a stress intensity factor that may be readily related to the work of adhesion. On the other hand, the tangential stress results in a (mode II-type) stress intensity factor for wave propagation}:
\begin{equation}
  \label{eqn:SIF}
  K_w \sim (\tau_r + 4G/\pi)\sqrt{x}
\end{equation}
in an appropriate singular limit. Here $x$ is the distance from the tip of the detachment zone as measured within the stationary/stick zone. \mt{A more general method for computing stress intensity factors such as Eq.~\ref{eqn:SIF} for the case $\nu \neq 0.5$, where closed form solutions are generally not obtainable, is presented in Ref.~\cite{AnsariViswanathan_PhysRevE_2022}. The consequences of Eq.~\ref{eqn:SIF} are discussed further in the next section}.

\subsection{Predicting the onset of stick--slip}
\label{subsec:phaseDig}

The occurrence of a stress singularity ahead of the detachment zone, Eq.~\ref{eqn:SIF}, represents a necessary condition for effecting wave motion. For a sufficient condition, reattachment of the detached interface must be considered in more detail. However, for the present, we contrast Eq.~\ref{eqn:SIF} with the analogous case of transition from static-to-dynamic friction outlined recently \cite{SvetlizkyETAL_PhysRevLett_2017}. According to this work, dynamic sliding occurs when
\begin{equation}
  \label{eqn:shearSIF}
  K_w \sim \tau_r\sqrt{x}
\end{equation}
reaches a critical value depending on the nature of the contact. The notation here is the same as that in Eq.~\ref{eqn:SIF}. While there has been much debate about the nature of shear cracks in general adhesive contacts (see for instance, Ref.~\cite{kendall2021energizing} and references therein), we nonetheless explore the consequences of Eq.~\ref{eqn:SIF} by comparing it with Eq.~\ref{eqn:shearSIF}, since the former naturally emerges as the solution of a rigorous boundary value problem.

Based on these two equations, a phase diagram may be developed showing when stick--slip motion via detachment waves (Eq.~\ref{eqn:SIF}, orange curve) should occur at the expense of uniform sliding (Eq.~\ref{eqn:shearSIF}, blue curve), see Fig.~\ref{fig:phaseDig}. The dashed curve denotes a limiting remote stress $\tau_r^*$ beyond which wave propagation does not occur. Likewise, $\Delta x_C$ is the critical detachment zone size below which wave propagation is not feasble.

\begin{figure}
  \centering
  \includegraphics[scale=0.8]{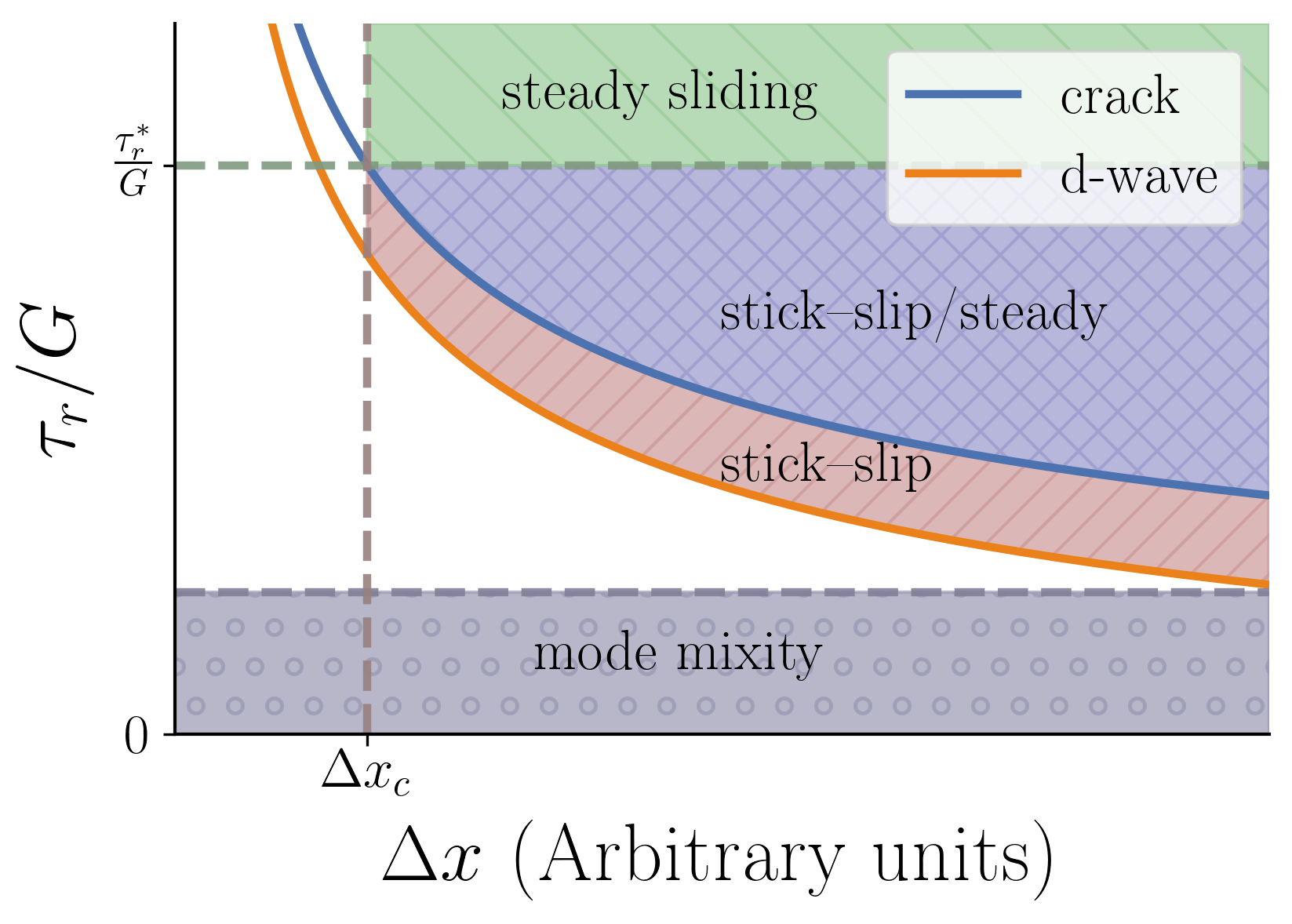}
  \caption{Qualitative phase diagram comparing Schallamach-type detachment waves with shear cracks to predict when stick--slip should occur at the expense of steady sliding.  The orange curve (d-wave) represents the equivalent stress intensity factor for a moving detachment wave ($\nu=0.5$) while the blue curve is for an equivalent shear crack of the same initial extent. The shaded regions are labeled based on which phenomenon, stick--slip or steady sliding/crack growth, is expected to occur.}
  \label{fig:phaseDig}
\end{figure}

The diagram is applicable to any material pair capable of forming adhesive contact (with one material significantly stiffer than the other) and must be interpreted as follows. Consider an adhesive interface that is loaded at constant remote $\sV$ as described in Sec.~\ref{sec:exptl}. If a detachment zone of size $\Delta x > \Delta x_c$ is nucleated, the total shear stress rises continuously since the interface remains stationary. When it crosses the orange curve, the necessary condition for wave motion is satisfied, however, the interface still remains stationary so that sliding does not yet start. When the remote $\tau_r$ then crosses the blue curve at the corresponding $\Delta x$ value, a crack begins to propagate at which point the entire interface can slip, or, if readhesion occurs, a single detachment wave propagates through the interface. If the latter possibility, the shear stress can reduce significantly to any point within the orange dashed region for continued propagation. The fact that wave motion occurs at constant speed $c_w$ implies that the shear stress settles at a fixed value within this orange zone very soon after nucleation.

On the other hand, if interface adhesion is not strong enough to effect reattachment, a single rupture will likely propagate through the interface, and cause in the manner discussed in Ref.~\cite{SvetlizkyETAL_PhysRevLett_2017}. Hence depending on the strength of adhesive interaction at the interface, either possibility---stick--slip or steady sliding---can occur within the blue (cross-hatched) zone. Once we cross the wave propagation threshold $\tau_r^*$, detachment waves are no longer possible so that only steady sliding results. The critical value $\Delta x_c$ is likewise determined by the intersection between the threshold line and the blue curve---below this nucleation width, wave propagation is no longer feasible.

\subsection{Exploiting the dislocation analogy}

An alternative theoretical development to the one discussed so far in this section is to exploit the dislocation analogy presented in Sec.~\ref{subsec:dislocations}. This can be used, for instance, to compute the shear force necessary for Schallamach wave propagation to occur using force expressions for elastic dislocations \cite{Nabarro_CrystalDislocations_1967}. The shear force $F_s$ required to move the wave must be balanced by the adhesion hysteresis in detachment and reattachment at the interface. Using this balance principle, we obtain the relevant critical shear force as
\begin{equation}
	F_s^c = \frac{a\,L_c\, \Delta W}{|\vec{b}|}
\end{equation}
where $\Delta W$ is the adhesion hysteresis, $L_c, a$ are the contact length and width, respectively. Using $|\vec{b}| = 255$ $\mu$m, $2a = 1$ mm, $L_c = 2.5$ cm for the data presented in Sec.~\ref{sec:exptl}, and $\Delta W \simeq 10$ mJ/m$^2$ for PDMS \cite{ChaudhuryETAL_JApplPhys_1996}, $F_s^c$ is estimated to be $0.5$ mN. This is the minimum force needed to propagate a single wave through the contact region. The wave pulse can thus travel through the interface at a much lower stress than that needed for nucleation, $F_c \simeq 1.6$ N for the data in Fig.~\ref{fig:ScW-SP-forces}. This thus explains the steep tangential force reduction after wave propagation begins---as the interface relaxes, the tangential force continues to decrease until it either equals $F_s^c$ or the wave exits the contact region.

Note that in the above, we've explicitly discussed a solitary Schallamach wave, but these considerations nonetheless apply to other Schallamach-type waves as well.

\section{Discussion}

Having presented both experimental and theoretical results on Schallamach waves and their properties, we now discuss several potential extensions and some unexpected implications of our work. 

\subsection{Noteworthy siblings of the Schallamach wave}

We have so far studied the underlying causes for nucleation---tensile necking, compressive buckling and compression-induced slip---of Schallamach-type waves. Given the generality of these mechanisms, one would not be hard pressed to expect similar waves at any deformable adhesive interface. One such example that has received some attention in the past is composite interfaces \cite{Kendall_Nature_1976, Kendall_CompInt_1996}. Here adhesion is externally induced (via a thin binder layer) and relative deformation between adjacent layers of the composite can nucleate Schallamach-type waves at the interface. Consequently, the intriguing possibility of these interfaces being toughened by such wave propagation events has been discussed extensively \cite{ Kendall_1978, Gittus_PhilMag_1975}.

In this context, we draw attention to three analogous waves in the world of soft-bodied locomotion \cite{GrayLissman_JExptBio_1938, Trueman_SoftBodiedLocomotion_1975, Gray_AnimalLocomotion_1968, LaiETAL_JExptBio_2010}. These organisms do not possess specialized limbs for effecting motion, so must rely on elastic deformations of their entire body structures to locomote \cite{Chapman_BiologicalReviews_1958}. They do so by propagating Schallamach-type waves along the length of the zone of contact between their body and the ground.

Direct waves---analogous to slip pulses---are compressive and used by some polychaete worms (e.g., \emph{Arenicola}) to effect motion within distinct \lq slip\rq\ segments propagating from the tail to the head \cite{Wells_QJMicSci_1954}. The equivalent of the separation or detachment pulse is the retrograde wave, which occurs in some Nemertines such as \emph{Rhynchodemus}. Here, a local tensile detachment zone is formed and propagates from the organism's head to its posterior end. Finally, the equivalent of the Schallamach wave is looping locomotion, seen in some \emph{Planaria}. Here, a detachment loop is formed at the rear end due to elastic buckling of body segments; the loop is then propagated towards the head, effecting slip as it moves \cite{Gray_AnimalLocomotion_1968}.

The fact that these mechanical equivalents in biology are the only three waves that have hitherto been recorded in the context of locomotion, begs the question of whether the three Schallamach--type waves are the only ones capable of mediating stick--slip motion at interfaces. The answer, as discussed in Fig.~\ref{fig:phaseDig_Expt} earlier, appears to be in the affirmative, atleast for low $V_0$ and normal loads. It is noteworthy that these biological analogues have also been considered as conceptual models, by one of the originators of the dislocation concept, Egon Orowan, in order to explain glide motion in metallic crystals \cite{Orowan_1954}.

\subsection{Viscoelastic and nonlinear effects on wave propagation}

Viscoelastic effects can have a significant influence on surface wave motion, yet incorporating them within the elastic framework of Sec.~\ref{subsec:continuum} remains quite challenging. In fact, even the elementary case of Rayleigh waves becomes remarkably complicated by the introduction of the simplest of linear viscoelastic constitutive relations \cite{CurrieETAL_1977}. Waves corresponding to other more complicated boundary conditions appear to be hopelessly intractable at present, barring a couple of grossly simplified examples \cite{BrenerETAL_EurPhysJE_2005, MartinsETAL_JVibrAcous_1995, Caroli_PhysRevE_2000}. This problem does warrant attention, given its importance from an experimental point of view---Schallamach wave properties are influenced by the materials' loss and storage moduli \cite{BarquinsCourtel_Wear_1975, RandCrosby_ApplPhysLett_2006} as well as other rate dependent properties \cite{RobertsThomas_1975, RobertsJackson_1975}. Consequently, rubber friction does obey the Williams-Landel-Ferry relation, typically associated with viscoelastic materials \cite{Grosch_1963}.

\subsection{Other nanoscale models for Schallamach-type waves}
\label{subsec:microscale}

While interface dynamics accompanying Schallamach waves is certainly describable at the continuum scale as discussed in Sec.~\ref{sec:theory}, a number of interesting phenomena arise when a general microscopic picture of interface friction is considered. Such a description for polymers was, unsurprisingly, put forth by Schallamach himself in a lesser known paper from 1963 \cite{Schallamach_Wear_1963}. This paper was focused on the dynamics of individual bonds at the interface, and did not consider collective effects between several bonds at once. The latter is the subject of a well-known model propounded by Prandtl and Tomlinson in the late 1920s \cite{Tomlinson_1929, Prandtl_1928}. 

Subsequently, several microscopic models have emerged, each treating bonds between the sliding bodies at the interface in a statistical manner. For instance, the simplest model consists of a network of horizontal and vertical springs that interconnect a set of rigid sliders driven by an external force. Models such as these show traveling solitary wave solutions that, at first glance, are remarkably reminiscent of Schallamach-type waves \cite{BraunPeyrard_PhysRevE_2012}. However, a closer look reveals that they do not explicitly account for large interface detachment nor do they result in interface slip. Yet, these models have found much favour within the community, likely because they are more easily amenable to incorporating viscous damping/dissipation effects and that their reduced dimensionality (most are quasi 1D) allows easier multi-scale analyses and numerical solution \cite{VanossiETAL_2013}. Complexity in this approach may be introduced by altering the nature of the springs (thereby the interaction) between the rigid sliders to more closely represent molecular interactions \cite{Filippov_PhysRevLett_2004, ChernyakLeonov_1986}. Unfortunately, just as with the continuum models of Sec.~\ref{sec:theory}, this often comes at the expense of tractability. Evaluating the consequences of these models constitutes an active area of research.

The generality of this microscopic approach reveals a fundamental kinship between the Prandtl-Tomlinson approach and two other famous discrete models---the Frenkel-Kontorova (FK) and Burridge-Knopoff (BK) models---that have their origins in crystal physics and tectonic fault dynamics \cite{FrenkelKontorova_1938, BurridgeKnopoff_1967}. Unlike the continuum elastic theories described in Sec.~\ref{sec:theory}, statistical models of the FK/BK type are inherently nonlinear. They consequently show very rich dynamics, such as traveling topological defects (kinks), localized breathers and even discrete effects \cite{BraunKivshar_1998, CartwrightETAL_PhysRevLett_1997, Muratov_1999}. Whether or not these solutions share similarities with Schallamach-type waves will only be revealed by additional experiments. We believe that the solitary Schallamach wave will play a key role in this regard.

The microscopic approach described in this section certainly has its merits, in that it can be applied to systems as diverse as tectonic faults and crystal lattices. Coupled with the experimental observations and the continuum--scale descriptions discussed in the previous section, interesting analogies between Schallamach-type waves and other traveling wave phenomena in these diverse systems begin to emerge.

Likewise, the addition of non-linearity---both material and geometric---to the governing equations results in the emergence of novel wave properties \cite{Whitham_1975}. When the governing wave equations are dispersive and non-linear, the possibility of the two effects exactly balancing results in the propagation of special particle--like waves called solitons \cite{Newell_1985}. In addition to the linear elastic waves discussed earlier, solitons can also retain their shape while propagating over long distances. However, the importance of nonlinear effects has not yet been firmly established and warrants additional detailed experiments.

\subsection{Exploiting the fracture and dislocation analogies for Schallamach waves}

 The wave-dislocation analogy raises the intriguing possibility of using Schallamach-type waves as a model to study dynamic dislocation behavior. For instance, one can contrive the use of interacting Schallamach waves to study time-dependent nanoscale phenomena such as dislocation coalescence. The existence of opposite moving detachment waves can then be used to \lq simulate\rq\ pairs of like and unlike dislocations. At present, questions such as these have only been treated using energetics \cite{Weertman_1966} or large scale molecular dynamics simulations with significant experimental uncertainty.

 Just as with the dislocation picture, the close fracture correspondence presented in Sec.~\ref{subsec:phaseDig} affords the possibility of using Schallamach-type  waves to study fracture problems and the transition from stick--slip to steady sliding. For instance, how do moving rupture fronts interact with crack pairs? One can also envisage the mechanical/frictional equivalent of the well-known Rice-Thomson model of dislocation emission from crack tips \cite{RiceThomson_1974}. If a moving rupture front delineating stationary and slipped zones propagates through the interface, can it become diffuse by interacting with a Schallamach wave? How does this behaviour change between the Schallamach wave and its dual? Finally, the length scale $\Delta x_c$ introduced in Fig.~\ref{fig:phaseDig} places bounds on the smallest Schallamach-type waves that may be observed in any system. For elastomers, we find that $\Delta x_c \sim 100$ nm which raises the question of whether nanoscale Schallamach waves can exist in any system. If so, detailed observations of their propagation properties could help explain stick--slip phenomena in a much wider variety of systems.

 We believe that these fascinating questions can significantly enhance our understanding of nanoscale phenomena using insights from seemingly unrelated Schallamach-type waves. We end by quoting Schallach once more \cite{Schallamach_Wear_1971}
\begin{displayquote}
  \textsl{The discussion of our experimental observations has been mostly conjecture, but may nevertheless suggest new approaches to the study of rubber friction.}
  \end{displayquote}
Little did Schallamach expect that his insights have helped open doors to linking several disparate fields together. We have every reason to believe that the legacy of Schallamach's discovery will endure for another fifty years.

\section{Summary and future outlook}

As we said at the beginning of this article, Schallamach's humble question has opened far reaching links between rubber friction and phenomena in domains as diverse as biological locomotion, nanoscale mechanics and interface fracture. Let us first attempt a summary of the salient features. We've seen how Schallamach waves nucleate due to a surface buckling instability, and how they subsequently propagate through the interface at speeds much lower than any elastic wave speed. Schallamach's observations, followed by several others, were all made using a spherical indenter geometry. Changing from this geometry to a cylindrical one has allowed the isolation of single or solitary Schallamach waves. Not only do these waves offer a clear quantitative picture of propagation dynamics, they also reveal remarkable similarities with propagating waves in other systems, e.g., biological locomotion. A second major advantage of the cylindrical geometry is that it enables observation of another slow moving detachment wave---the separation pulse or the dual of the Schallamach wave. Together, the Schallamach wave and its dual represent the only two possibile mechanisms for waves of detachment at soft adhesive interfaces. A final detachment-less wave is also observed---the slip pulse---which completed our trio of Schallamach-type waves. We noted that there are exact analogues of these Schallamach--type waves in biological locomotion. Based on the two key features of Schallamach--type waves, \emph{viz.} that they move at slow speed and result in constant slip in thier wake, we explored possible continuum theories for explaining their propagation. The failure of conventional elastodynamic models to explain both key features was remedied---for the Schallamach wave and its dual atleast---by the use of a model-free mixed boundary value problem framework. This approach has helped bridge the gap between theory and experiment, but left several questions unanswered. Primarily among these is the role of viscoelasticity---that appears to completely obliterate any signs of a closed-form solution. There remains much to be done here to explain the role of key phenomena such as the observed Williams-Landel-Ferry relationship in rubber friction as well as the origin of contact relaxation effects.

Moving onto smaller scales, we deliberated on the rather intimate link between friction phenomena in general and nanoscale mechanics. Two noteworthy examples are the correspondence between Schallamach-type waves and interface crack pairs, and the dislocation analogy. This then raises the natural question of whether Schallamach-type waves can be used to study the dynamics of nanoscale mechanics and fracture phenomena. It is presently standard practice to treat these problems using energetics, so that even rudimentary understanding of their time-dependence has remained elusive. Little did Schallamach know that the waves that are now named after him, could be the bridge that connects such diverse areas. There is much yet to be unearthed in this matter and we hope that a reader of this article will fire the first shots.

\section*{Acknowledgements}
We would like to thank Professor Kevin Kendall, FRS, for his careful reading of a draft version of this paper, and for his constructive comments and criticisms. This work was supported in part by a grant from the Indian Institute of Science, Bangalore (KV) and NSF awards DMR 2104745 and CMMI 2100568 (SC)

\appendix

\section*{Appendix: A primer on elastic surface waves}
\label{sec:appendix}
The dynamic behaviour of isotropic linear elastic materials is determined by two characteristic material velocities---the shear $c_T$ and longitudinal $c_L$ wave speeds. The displacement field $\vec{u}$ in an elastic body is determined as a suitable linear combination of an irrotational component $\vec{u}_L$ and an incompressible component $\vec{u}_T$, each obeying a linear wave equation \cite{Achenbach_WavePropagation_2012}
\begin{align}
  \label{eqn:waveEqn}
  \pd{\vec{u}_L}{t^2} &= c_L^2 \nabla^2 \vec{u}_L\quad\quad\quad \nabla \times \vec{u}_L = 0 \\\notag
  \pd{\vec{u}_T}{t^2} &= c_T^2 \nabla^2 \vec{u}_T\quad\quad\quad \nabla \cdot \vec{u}_T = 0
\end{align}
These equations must be supplemented by boundary conditions (BCs) and initial conditions (ICs) pertinent to the particular physical problem. The BCs could include specification of either the stresses/displacements or both on part of or the entire surface of the elastic body. Unlike surface waves, longitudinal and transverse waves desribed by $\vec{u}_L, \vec{u}_T$ travel in the bulk of the material (infinite space). Notably, $\vec{u}_L, \vec{u}_T$ waves are non-dispersive---their propagation speed does not depend on the wavelength. Specifically, linear, non-dispersive waves can propagate without changing their shape whatsoever \cite{Whitham_1975}. PDEs such as Eq.~\ref{eqn:waveEqn} govern only wave propagation and say nothing about how or why they must nucleate in the first place. For the elastic polymer in the experiments of Sec.~\ref{sec:exptl}, the wave speeds $c_L, c_T \sim 100$ m/s.

The waves that can then propagate at or near an interface or free surface are determined by specific boundary conditions. The simplest surface wave is the Rayleigh wave and it occurs on a stress free surface of an elastic solid \cite{Strutt_ProcLondMathSoc_1885}. Consider a semi-infinite elastic solid $z\leq 0$ with the $z=0$ plane as its bounding surface. These waves are exponentially damped into the solid $z<0$ and propagate on the surface $z=0$ at a fixed velocity $c_R < c_T < c_L$, that is determined by the material's Poisson's ratio. Rayleigh waves are well known free surface waves that usually accompany earthquakes; they arrive after the first acoustic signals (since $c_R < c_T, c_L$) and are non--dispersive. 

In contrast, when two elastic materials are brought into contact, a Rayleigh wave cannot propagate at the interface, because the surface is no longer stress--free. On the contrary, if the two solids have continuous displacements and stresses at the interface (so-called \lq welded contact\rq ), a Stoneley wave can propagate instead \cite{Stoneley_ProcRoySocA_1924}. The existence of Stoneley waves is tied to the elastic properties of the two contacting materials. However, when they do exist, these waves propagate at a fixed velocity, which is a function of the two sets of elastic properties. Just like Rayleigh waves, Stoneley waves are also non--dispersive.

If the \lq welded\rq\ contact described above is replaced by a frictionless contact $(\sigma_{xz} = \sigma_{yz} = 0$), then an Achenbach--Epstein (AE) wave or generalized Rayleigh wave can propagate instead of Stoneley waves \cite{AchenbachEpstein_JEnggMechDiv_1967}. The requisite boundary conditions now involve continuous displacements across the interface, equal normal forces and zero tangential forces on both bodies. The existence of AE waves at an interface between two solids depends on the elastic properties of the two materials. Like both Rayleigh and Stoneley waves, AE waves also propagate at a fixed velocity, dependent on the two sets of material elastic properties, and are non--dispersive.

\bibliography{bibfile}
\bibliographystyle{vancouver}
\end{document}